# Effect of surface ionic screening on polarization reversal scenario in ferroelectric thin films: crossover from ferroionic to antiferroionic states

Anna N. Morozovska[1,2,*], Eugene A. Eliseev[3], Anatolii I. Kurchak[4], Nicholas V. Morozovsky[1], Rama K. Vasudevan[5], Maksym V. Strikha[4,6], and Sergei V. Kalinin[5,†]

[1] *Institute of Physics, National Academy of Sciences of Ukraine,*
*pr. Nauky 46, 03028 Kyiv, Ukraine*

[2] *Bogolyubov Institute for Theoretical Physics, National Academy of Sciences of Ukraine,*
*14-b Metrolohichna str. 03680 Kyiv, Ukraine*

[3] *Institute for Problems of Materials Science, National Academy of Sciences of Ukraine, Krjijanovskogo 3, 03142 Kyiv, Ukraine*

[4] *V.Lashkariov Institute of Semiconductor Physics, National Academy of Sciences of Ukraine,*
*pr. Nauky 41, 03028 Kyiv, Ukraine*

[5] *The Center for Nanophase Materials Sciences, Oak Ridge National Laboratory,*
*Oak Ridge, TN 37831*

[6] *Taras Shevchenko Kyiv National University, Radiophysical Faculty*
*pr. Akademika Hlushkova 4g, 03022 Kyiv, Ukraine*

**Abstract**

Nonlinear electrostatic interaction between the surface ions of electrochemical nature and ferroelectric dipoles gives rise to the coupled ferroionic states in nanoscale ferroelectrics. Here, we investigated the role of the surface ions formation energy value on the polarization states and polarization reversal mechanisms, domain structure and corresponding phase diagrams of ferroelectric thin films. Using 3D finite elements modeling we analyze the distribution and hysteresis loops of ferroelectric polarization and ionic charge, and dynamics of the domain states. These calculations performed over large parameter space delineate the regions of single- and poly- domain ferroelectric, ferroionic, antiferroionic and non-ferroelectric states as a function of surface ions formation energy, film thickness, applied voltage and temperature. We further map the analytical theory for 1D system onto effective Landau-Ginzburg free energy and establish the correspondence between the 3D numerical and 1D analytical results. This approach allows performing the overview of the phase diagrams of ferroionic systems and exploring the specific of switching and domain evolution phenomena.

[*] corresponding author, e-mail: anna.n.morozovska@gmail.com
[†] corresponding author, e-mail: sergei2@ornl.gov

# I. INTRODUCTION

Ferroelectric phase stability requires effective screening of the polarization bound charge at surfaces and interfaces with non-zero normal component of polarization [1, 2, 3]. Traditionally, the theoretical analysis of bulk ferroelectric state assumes either complete screening of polarization by (effective) electrodes, or considers the emergence of the multidomain states as a pathway to minimize depolarization energy [1-3]. Rapid growth of thin-film applications of ferroelectrics in 90ies necessitated analysis of the microscopic mechanisms active at ferroelectric interfaces, preponderantly effects stemming from non-zero space separation between polarization and screening charges [4, 5, 6, 7, 8]. These effects are often introduced via the dead layer [2] or physical gap [9] concepts, postulating the presence of the thin non-ferroelectric layer or gap separating ferroelectric and metal and have been shown to agree with *ab initio* calculations [10, 11]. Because of the long-range nature of the depolarization effects the incomplete surface screening of ferroelectric polarization leads to pronounced finite size effects [12] and nontrivial domain structure dynamics [1] in the presence of thin dead layers and gaps. These in turn causes unusual phenomena near the electrically opened surfaces such as correlated polarization switching, formation of flux-closure domains in multiaxial ferroelectrics [13, 14, 15, 16], domain wall broadening in uniaxial [17, 18] and multiaxial ferroelectrics [15, 16]. Further examples of these behaviours include the crossover between different screening regimes in ferroelectric films [19, 20] and p-n junctions induced in 2D-semiconductors [21] separated by the ultra-thin gap from the moving domain wall - surface junctions.

However, the dead layer approximation completely ignores many characteristic details of the thermodynamics of screening process. In particular, the stabilization ferroelectric state in ultrathin perovskite films can take place due to the chemical switching (see e.g. [22, 23, 24]), and the screening via ionic adsorption is intrinsically coupled to the thermodynamic of surface electrochemical processes. This coupling results in non-trivial effects on ferroelectric phase stability and phase diagrams [25, 26]. Similar effects are expected for screening by electrodes with finite density of states [27]. However, the surface state of materials in contact with atmosphere is usually poorly defined, due to the presence of mobile electrochemically active and physically sorbable components in ambient environment [28, 29]. The effect of adsorption of oxygen and hydrogen on the work function, reversible polarization value, dipole moment of the unit cell and free energy of the semiconductor ferroelectrics had been investigated experimentally and theoretically [30, 31, 32, 33], albeit total volume of research in this area is extremely small. However, no comprehensive or general approaches have been developed.

The early theoretical analyses, though studied the properties of ferroelectric material in detail, typically ignored the nonlinear tunable characteristics of surface screening charges [30 – 33]. A complementary thermodynamic approach was developed by Stephenson and Highland (**SH**) [25, 26], who consider an ultrathin film in equilibrium with a chemical environment that supplies ionic species (positive and negative) to compensate its polarization bound charge at the surface.

Recently, we modified the SH approach allowing for the presence of the gap between the ferroelectric surface covered by ions and the SPM tip [34, 35, 36, 37], and developed the analytic description for thermodynamics and kinetics of these systems. The analysis [34 – 36] leads to the elucidation of the ferroionic states, which are the result of nonlinear electrostatic interaction between the surface ions with the charge density obeyed Langmuir adsorption isotherm and **single-domain** ferroelectric polarization. The properties of ferroionic states were described by the system of coupled equation.

Here, we study the stability of ferroionic states with respect to the **domain structure formation** and **polarization reversal scenarios** in thin ferroelectric films covered by ions. Our results, presented below, reveal unusual dependences of the film polar state and domain structure properties on the ion formation energies and their difference, and, even more, unexpected, on the applied voltage. That say one can expect to be faced with the electric field-induced phase transitions into ferroionic state in thin films covered with ion layer of electrochemically active nature.

The manuscript is structured as following. Basic equations with boundary conditions are given in **section II**. Numerical results presented in **section III** demonstrate the effect of surface ions formation energy on stable polarization states, evolution of domain structure and surface charge during polarization reversal. To get insight into numerical results we presented simplified analytical modeling of ferro-ionic system behavior based on the free energy with renormalized coefficients in **section IV.** Distinctive features of polarization reversal in thin ferroelectric films covered by ions are discussed in **section V**. Electrostatic problem and derivation of the free energy with renormalized coefficients are given in **Appendixes A** and **B**, respectively. Parameters used in calculations and auxiliary figures are listed in **Supp. Mat**.[38].

## II BASIC EQUATIONS WITH BOUNDARY CONDITIONS

Here we consider the system consisting of electron conducting bottom electrode, ferroelectric (FE) film, surface ions layer with charge density $\sigma(\varphi)$, ultra-thin gap separated film surface and the top electrode (either ion conductive planar electrode or flatted apex of SPM tip) providing direct ion exchange with ambient media, as shown in **Fig. 1(a).** Mathematical statement of the problem is listed in Refs.[35 – 36] as well as in **Appendix A.**

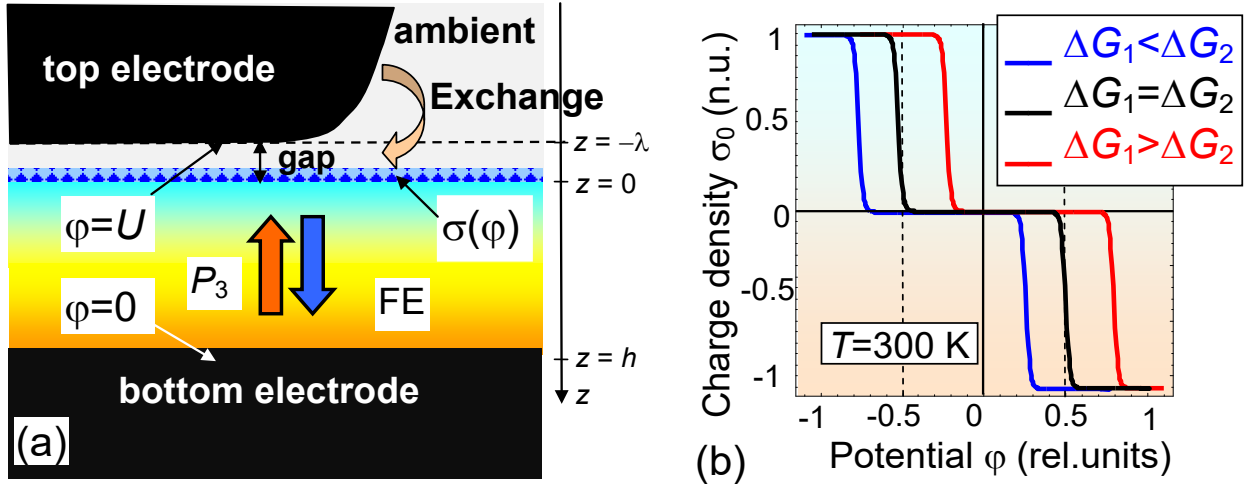

**FIGURE 1**. **(a)** Layout of the considered system, consisting of electron conducting bottom electrode, ferroelectric (FE) film, surface ions layer with charge density $\sigma(\varphi)$, ultra-thin gap separated film surface and the top electrode providing direct ion exchange with ambient media (from bottom to the top). **(b)** Schematic dependence of the equilibrium surface charge density $\sigma_0$ vs. acting potential $\varphi$ calculated for $\Delta G_1^{00} < \Delta G_2^{00}$ (blue curve), $\Delta G_1^{00} = \Delta G_2^{00}$ (black curve) and $\Delta G_1^{00} > \Delta G_2^{00}$ (red curve) (Adapted from Ref. [35]).

We account for the presence of the ultra-thin dielectric gap between the probe tip and the ferroelectric surface. At that, the linear equation of state $\mathbf{D} = \varepsilon_0 \varepsilon_d \mathbf{E}$ relates the electrical displacement $\mathbf{D}$ and electric field $\mathbf{E}$ in the gap ($\varepsilon_0$ is a universal dielectric constant and $\varepsilon_d \sim 1$ is the relative permittivity of the physical gap media (vacuum, air or inert gas environment)). A wide band-gap ferroelectric film can be considered dielectric. Quasi-static electric field inside the ferroelectric film $\varphi_f$ satisfies electrostatic equations for each of the medium (gap and ferroelectric film). The boundary conditions (**BCs**) for the system are the equivalence of the electric potential to the applied voltage $U$ at the top electrode (or SPM tip apex modeled by the flat region $z = -\lambda$) and the equivalence of the normal component of electric displacements to the ionic surface charge density $\sigma[\varphi(\vec{r})]$ at $z = 0$; the continuity of the electric potential and normal component of displacements $D_3 = \varepsilon_0 E_3 + P_3$ and $D_3 = \varepsilon_0 \varepsilon_d E_3$ at gap - ferroelectric interface $z = 0$; and zero potential at the bottom conducting electrode $z = h$ [see **Fig. 1**].

The polarization components of uniaxial ferroelectric film depend on the inner field $\mathbf{E}$ as $P_1 = \varepsilon_0(\varepsilon_{11}^f - 1)E_1$, $P_2 = \varepsilon_0(\varepsilon_{11}^f - 1)E_2$ and $P_3(\mathbf{r}, E_3) = P_3^f(\mathbf{r}, E_3) + \varepsilon_0(\varepsilon_{33}^b - 1)E_3$, where background permittivity $\varepsilon_{33}^b \leq 10$ [39]. The dielectric permittivity $\varepsilon_{33}^f$ is related with the ferroelectric polarization $P_3$ via the soft mode. The evolution and spatial distribution of the ferroelectric polarization $P_3^f$ (further abbreviated as $P_3$) is given by the time-dependent LGD equation:

$$\Gamma \frac{\partial P_3}{\partial t} + \alpha P_3 + \beta P_3^3 + \gamma P_3^5 - g \frac{\partial^2 P_3}{\partial z^2} = E_3, \qquad (1)$$

In Eq.(1), the kinetic coefficient $\Gamma$ is defined by phonon relaxation; $\alpha = \alpha_T(T_C - T)$, $\beta$ and $\gamma \geq 0$ are the coefficients of LGD potential $F(P_i, X_{ij}, \sigma)$ expansion on the higher polarization powers [40]; $T$ is the absolute temperature, $T_C$ is Curie temperature. The **BCs** for polarization at the film surfaces z = 0 and z = h are of the third kind $\left( P_3 \mp \Lambda_\mp \frac{\partial P_3}{\partial z} \right)\bigg|_{z=0,h} = 0$ and include extrapolation lengths $\Lambda_\pm$ [41, 42].

Equation for the surface charge is analogous to the Langmuir adsorption isotherm used in interfacial electrochemistry for adsorption onto a conducting electrode exposed to ions in a solution [43]. To describe the dynamics of surface ions, we use a linear relaxation model for their charge density, $\tau \frac{\partial \sigma}{\partial t} + \sigma = \sigma_0[\varphi]$, where the dependence of equilibrium charge density $\sigma_0[\varphi]$ on electric potential $\varphi$ is controlled by the concentration of surface ions $\theta_i(\varphi)$ at the interface z = 0 in a self-consistent manner [25, 26]:

$$\sigma_0[\varphi] = \sum_i \frac{eZ_i \theta_i(\varphi)}{A_i} \equiv \sum_i \frac{eZ_i}{A_i} \left( 1 + \left( \frac{p_{atm}}{p_{exc}} \right)^{1/n_i} \exp\left( \frac{\Delta G_i^{00} + eZ_i\varphi}{k_B T} \right) \right)^{-1}, \quad (2)$$

where $e$ is an elementary charge, $Z_i$ is the charge of the surface ions/electrons, $1/A_i$ is the saturation densities of the surface ions, at that $i \geq 2$ to reach the charge compensation, $p_{exc}$ is the partial pressure of ambient gas relative to atmospheric pressure $p_{atm}$, $n_i$ is the number of surface ions created per gas molecule, $\Delta G_i^{00}$ is the standard free energy of the surface ion formation at $p_{exc} = 1$ bar and $U = 0$. Schematic step-like dependence of the surface charge density $\sigma_0$ on the potential $\varphi$ is shown in **Fig. 1(b)**. The surface charge is small or zero at $\varphi = 0$, and then is abruptly increases for potential values $\varphi_i \approx -\Delta G_i^{00}/(eZ_i)$. The abrupt step-like dependence is left-shifted with respect to $\varphi = 0$ for $\Delta G_1^{00} < \Delta G_2^{00}$ (blue curve), almost symmetric for $\Delta G_1^{00} \approx \Delta G_2^{00}$ (black curve) and right-shifted for $\Delta G_1^{00} > \Delta G_2^{00}$ (red curve). For equal $\Delta G_1^{00} = \Delta G_2^{00} \equiv \Delta G$ the Langmuir isotherm is the even function of $\Delta G$, $\sigma_0(\varphi, \Delta G) \equiv \sigma_0(\varphi, -\Delta G)$ provided that other parameters correspond to the charge neutrality at zero potential.

Notably, the developed solutions are insensitive to the specific details of the charge compensation process [44], and are sensitive only to the thermodynamic parameters of corresponding reactions [45].

### III. EFFECT OF SURFACE IONS FORMATION ENERGY ON POLARIZATION REVERSAL AND DOMAIN STRUCTURE. FEM RESULS

Using 3D finite element modeling (FEM) we studied the polarization states in the ferro-ionic system. Following the terminology introduced in Refs.[34 – 36] we classified the possible polar state in the ferro-ionic system based on the free energy minima and hysteresis loops shaping. A general property of the ferroelectric film covered with ions is that it can be in the single-domain (**SD**) or poly-domain (**PD**) ferroelectric (**FE**), ferroionic (**FI**), antiferroionic (**AFI**) and ionic non-ferroelectric (**NFE**) equilibrium states depending on the applied voltage $U$, temperature T, film thickness h and ion formation energies $\Delta G_i^{00}$ (see **Figs 2**).

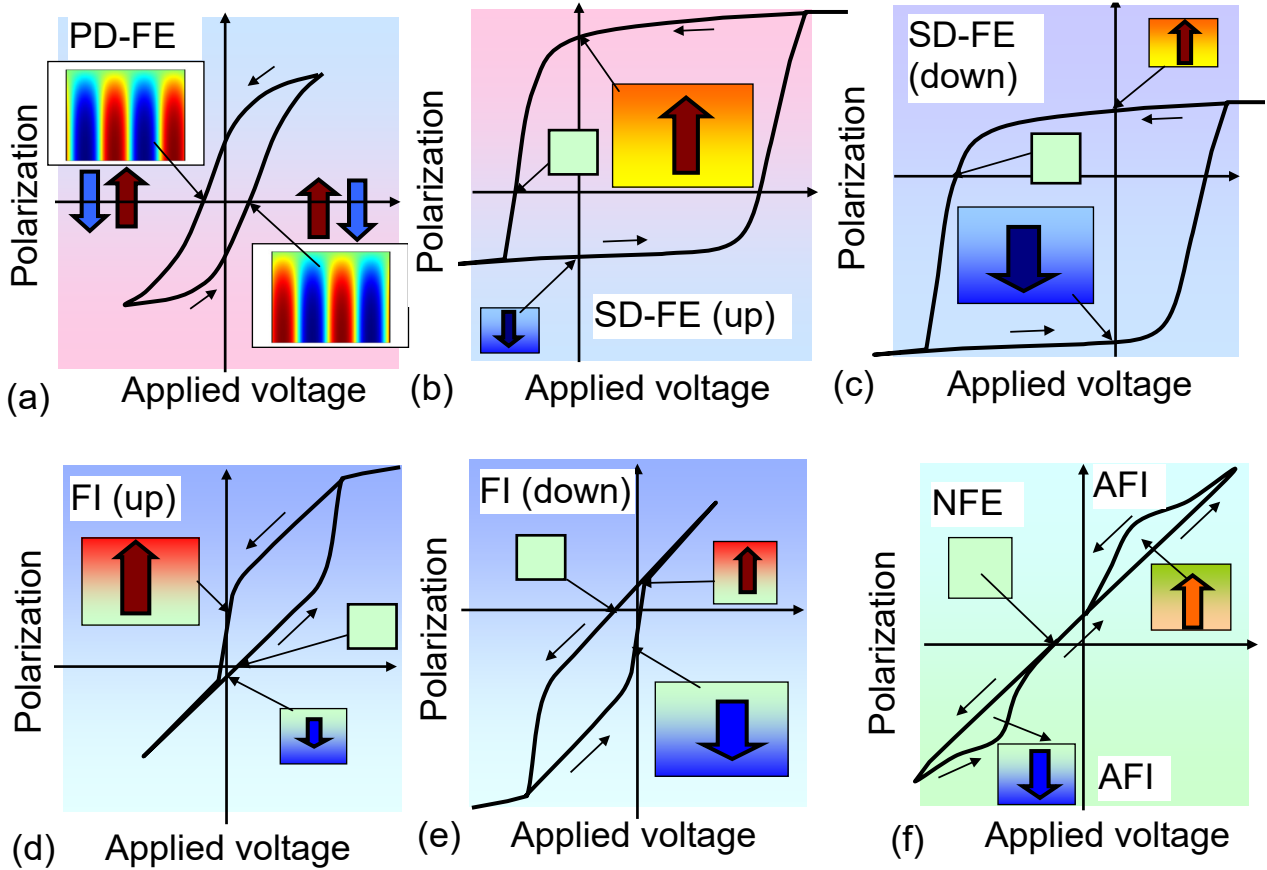

**FIGURE 2. (a)-(c)** Polar states classification in the ferro-ionic system based on the polarization distribution and hysteresis loop form. Characteristic hysteresis loops for poly-domain ferroelectric (**PD-FE**) state **(a),** positive **(b)** and negative **(c)** single-domain ferroelectric (**SD-FE**) states with dominant "upward" and "downward" polarizations, positive **(d)** and negative **(e)** ferroionic (**FI**) states with almost "upward" and "downward" polarizations; and **(f)** mixed antiferroionic (**AFI**) and ionic non-ferroelectric (**NFE**) state are shown. Polarization distribution corresponding to definite points at the hysteresis loop is shown by insets with upward or downward oriented arrows for a homogeneously polarized film, two upward and downward oriented arrows for a poly-domain state, and empty square for a zero homogeneous polarization.

The **FE** state is defined as the state with robust ferroelectric hysteresis between at least two (or more) absolutely stable ferroelectric polarizations, which have "positive" or "negative" projection with respect to the film surface normal [see **Figs. 2(a)-(c)**]. Due to the nonlinear voltage dependence of the surface ions charge density [Eq.(2)], the polarization screening by the ions result in one or more unstable positive or negative polarization states. FE states exist at film thickness above the critical value, $h \geq h_{cr}(T)$, at fixed temperature $T < T_C$. Usually the value $h_{cr}(T)$ depends on the screening conditions [1, 15, 46]. Due to the presence of the gap between the film surface and the tip electrode, FE states can be poly-domain (**PD-FE**) with a rather thin and tilted hysteresis loop at small amplitudes of applied voltage [see **Figs. 2(a)**]. Also FE states can be almost single-domain (**SD-FE**) with symmetric, dominantly positive or negative polarizations and correspond to a wide square-like hysteresis loop at higher voltage amplitudes [see **Figs. 2(b)** and **2(c)**]. Noteworthy SD-FE states can be more stable than the PD-FE ones for $\Delta G_1^{00} \neq \Delta G_2^{00}$, because positive and negative orientations of polarization are energetically equivalent only at zero applied voltage $U = 0$. The physical origin of SD-FE states voltage asymmetry is the electric field $E_{eff} \sim (\Delta G_1^{00} - \Delta G_2^{00})/h$ produced by surface ions with the charge density σ(U). The field $E_{eff}$ was introduced earlier [34 - 36] and its analytical form will be discussed in the next section. When this field of electrochemical nature is above certain critical value, SD-FE state becomes absolutely stable, while PD-FE state becomes metastable.

The **FI** state appears with increase of $E_{eff}$ as supported by the voltage bistability of ionic states and its boundary with SD-FE stable is conditional (resembling relative definition of ferroelectric and pyroelectric). FI state is characterized by a strongly asymmetric, horizontally shifted or vertically imprinted (or "minor") hysteresis loop between two polarization states (one is stable and another is metastable). At that the polarization orientation in the states can be the same (e.g. positive or negative in both states), and the values should be different (e.g. one big and another small) [compare **Fig. 2(d)** and **2(e)**]. FI states appear because one of polarization orientations loses its stability when the film thickness $h$ becomes less than the critical value $h_{cr}(T)$, and can exist for significantly different values of $\Delta G_i^{00}$ (e.g. for $|\Delta G_1^{00} - \Delta G_2^{00}| \geq 0.2$ eV) even in ultrathin films ($h \leq 2$ nm) at high temperatures. [2].

**NFE** state follows the FI state under the temperature increase or further decrease of the film thickness well below the critical value $h_{cr}(T)$. NFE state appears when two metastable polarization orientations [shown in **Fig. 1(f)** by upward and downward arrows] disappear at the energy relief. NFE state has no hysteresis properties at $U = 0$ in the thermodynamic limit and reveal electret-like polarization state $P_3 \sim E_{eff}(\sigma)$ induced by the field $E_{eff}$ [see **Fig. 2(f)**].

A double hysteresis loop opening appeared at high applied voltages $U$ corresponds to the novel antiferroionic (**AFI**) state. Since the energies of upward and downward polarization orientations gradually reach each other with the temperature increase the diffuse boundary between AFI and FI states is defined by the electrochemical properties of surface ions, in particular by the ion formation energies $\Delta G_i^{00}$.

Stable polarization states were calculated numerically in dependence on the surface ions formation energies $\Delta G_i^{00}$ at zero applied voltage ($U = 0$). Corresponding diagrams are shown in **Figs. 3(a), (b)** and **(c)** for 100-nm, 50-nm and 10-nm PZT films, respectively. Here, the calculations are performed for permittivity of the dielectric gap $\varepsilon_d = 1$, its thickness $\lambda = 0.4$ nm and saturation area of the surface ions $A_1 = A_2 = 10^{-18}$ m², other parameters are listed in **Table SI** in **Suppl. Mat** [38].

Green circles in the diagrams **Figs. 3(a)-(b)** correspond to the PD-FE state, blue and orange circles are negative and positive SD-FE states, which continuously transform into FI states. Dashed curves correspond to the contours of constant intrinsic electric field $E_{eff}$ induced by ions. Solid curves are drawn by eye. Note the that the domain states in the regions between dashed and the solid curves [some of the circles are semi-colored in **Fig. 3(a)-(b)**] are very sensitive to the initial seeding (random polarization distribution at $t=0$), and the film in the average (even being poly-domain) can have a nonzero unipolarity degree, i.e. the fraction (sizes and/or number) of domains with negative polarization orientation is not equal to the fraction with positive one.

A comparison of the polarization states diagram of 100-nm film [**Fig. 3(a)**] with the one of 50-nm film [**Fig. 3(b)**] shows that the region of PD-FE states (green circles between the solid curves) that extends from the diagonal $\Delta G_1^{00} = \Delta G_2^{00}$ becomes much thinner and smaller with the film thickness decrease. For 50-nm film the region narrows to the diagonal at $\Delta G_1^{00} = \Delta G_2^{00} \leq 0.7$ eV and only then slightly enlarges at $0.7 < \Delta G_i^{00} \leq 1$ eV. The regions of positive and negative SD-FE states, which are induced by the intrinsic field $E_{eff}$, expand with the increase of the difference $\left|\Delta G_1^{00} - \Delta G_2^{00}\right|$. Also the region of SD-FE state significantly increases with the film thickness $h$ decrease because $E_{eff} \sim \left(\Delta G_1^{00} - \Delta G_2^{00}\right)/h$. Note that the regions of FI state are located far from the diagonal at $\left|\Delta G_1^{00} - \Delta G_2^{00}\right| \geq 0.8$ eV and slightly expand with $h$ increase.

The diagram for the thinnest 10-nm film is very different from the ones for 50- and 100-nm films [see **Fig. 3(a)**]. FE states are absent for the case, because 10-nm film is paraelectric at $\sigma = 0$. Light green, blue and orange circles in the diagram **(c)** correspond to the electret-like NFE states, and FI states with negative and positive polarizations, respectively. The film is in a paraelectric state along the diagonal, $\Delta G_1^{00} = \Delta G_2^{00}$. A wide almost square region of NFE states corresponds to $\Delta G_i^{00} \geq 0.2$ eV. Note that NFE states in the regions between dashed and solid lines are very sensitive to the mesh details, and the film on the average (being in a NFE state) has a non-zero average polarization [all of the circles are semi-colored in **Fig. 3(c)**]. Alternatively, NFE states can be metastable (or slowly relaxing one), the dynamics difficult to capture via numerical model. Stable "up" and "down" FI states are caused by $E_{eff}$ increase and located in the two rectangles with rounded edges, $0 \leq \Delta G_i^{00} \leq 0.15$ eV. Note that the unusual physical picture shown in **Fig. 3(c)** corresponds to the stable polarization states at zero voltage $U = 0$.

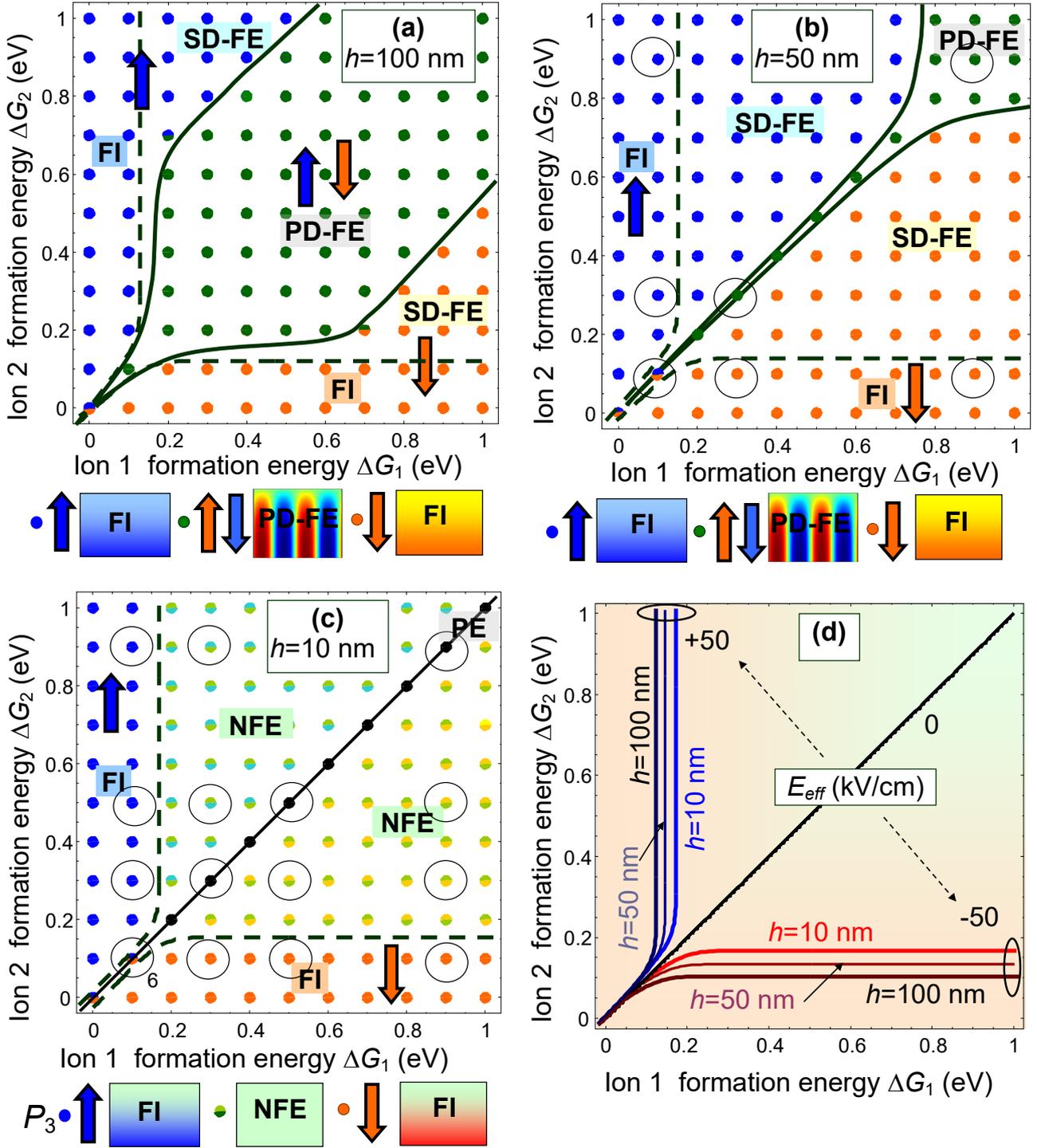

**FIGURE 3.** Polarization states in dependence on the surface ions formation energies $\Delta G_i^{00}$ calculated for PZT film thickness $h$ = 100 nm **(a)**; 50 nm **(b)** and 10 nm **(c)**. In the diagrams **(a, b),** dark green circles correspond to the poly-domain ferroelectric state (**PD-FE**), single-domain ferroelectric (**SD-FE**) states with opposite (negative and positive) polarization orientations are represented by blue and orange circles, respectively. SD-FE states continuously transform into ferroionic (**FI**) states (represented by blue and orange circles). In the diagram **(c),** black, light yellow-green, blue and orange filled circles correspond to the paraelectric-like (**PE**) state, electret-like nonferroelectric (**NFE**) states and ferroionic (**FI**) states with opposite polarization directions, respectively. Solid curves are drawn by eye. **(d)** Effective intrinsic field $E_0[\Delta G_i^{00}, h]$ (in kV/cm) depending on the surface ions formation energies $\Delta G_i^{00}$ calculated for PZT film thickness $h$ = 10, 50 and 100 nm (legends at the curves) at $T$ =

300 K; other parameters are listed in **Table SI.** Empty rings indicate the points, where the dependences $P(U)$ are shown in **Fig. 4** and **Figs. S1-S4**.

The voltage dependences of polarization $P(U)$ and surface charge $\sigma(U)$ were calculated at amplitudes (20 V) of applied voltages for a 10-nm PZT film (corresponding applied field is 2 GV/m). Results for $P(U)$ and $\sigma(U)$ are shown in **Fig. 4(a)** and **Fig. 4(b)**, respectively. These graphs indicate that symmetric, slightly or strongly asymmetric, and double hysteresis loops with pronounced coercive voltage (about 10 V) exist at $\Delta G_i^{00} \leq 0.5\text{eV}$. It is worthy of note that the spontaneous ferroelectric polarization disappears at zero voltage for $\Delta G_i^{00} > 0.2\text{eV}$ [as it is shown in **Fig. 3(c)**].

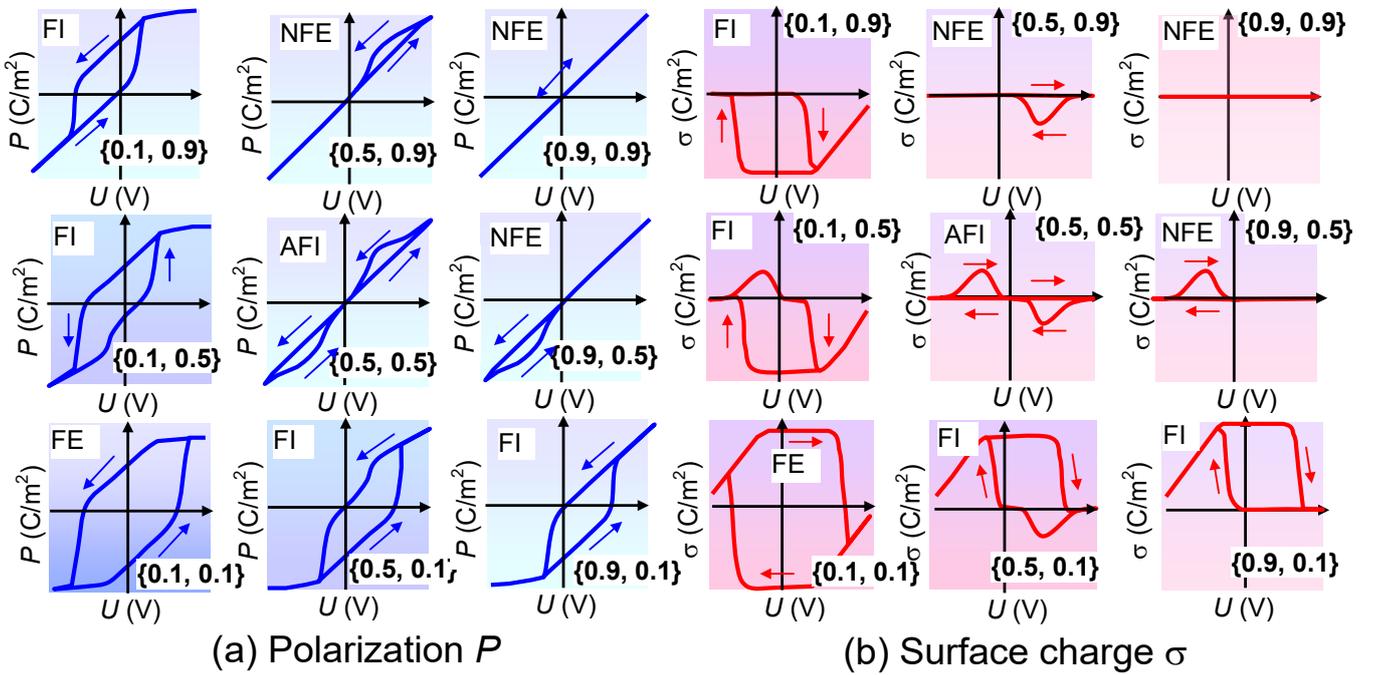

**FIGURE 4.** Dependences of average polarization $P(U)$ **(a)** and surface charge $\sigma(U)$ **(b)** on applied voltage U calculated at fixed values $\{\Delta G_1^{00}, \Delta G_2^{00}\}$ indicated in the labels (in eV) for a 10-nm PZT film. The voltage varies from −20 to +20 V. Polarization varies from −0.45C/m² to +0.45C/m². Surface charge varies from −0.35C/m² to +0.35C/m². Arrows indicate the direction of the loops pass-tracing. Other parameters are the same as in **Fig. 2.**

Comparison of the loops shown in **Fig. 4(a)** with the points on the diagram **Fig. 3(c)** corresponding to the same pair of ion formation energies $\{\Delta G_1^{00}, \Delta G_2^{00}\}$ leads to the following conclusions. If the film is in a FI state at $U = 0$, it remains in the state with the voltage amplitude increase, at that the polarization reversal along asymmetric hysteresis loop follows the single-domain scenario. When the film is in a NFE state at $U = 0$ it can remains NFE with the voltage increase at $\left| \Delta G_1^{00} - \Delta G_2^{00} \right| \leq 0.1\,\text{eV}$. However if the difference increases, $\left| \Delta G_1^{00} - \Delta G_2^{00} \right| > 0.2\,\text{eV}$, the film in NFE

state unexpectedly becomes FI with the voltage increase [see **Figs. S1-S2**]. Hysteresis loops of $P(U)$ and $\sigma(U)$ calculated for more pairs of $\{\Delta G_1^{00}, \Delta G_2^{00}\}$ and $h$ = 10 nm are shown in **Figs. S1-S2**. From the figures one can see that almost symmetrical hysteresis loops of $P(U)$ and $\sigma(U)$ appear at voltages amplitudes greater than 10 V and equal ion formation energies $\Delta G_1^{00} = \Delta G_2^{00} \leq 0.4$ eV.

At equal $\Delta G_1^{00} = \Delta G_2^{00} = 0.5$ eV the double hysteresis loop occurs indicating the transition to AFI states. The double loops degrade and eventually disappear with increasing the value of $\Delta G_1^{00} = \Delta G_2^{00}$ [see plots in **Figs. 4** for $\Delta G_1^{00} = \Delta G_2^{00} = 0.9$ eV]. The physical reason for abovementioned crossover from the single to double hysteresis loops is the voltage dependence of ionic charge magnitude [see **Fig. 1(b)**]. Here, the magnitude of ion charge increases, because its density depends exponentially on the voltage [see Eq.(2)]. This inevitably means that the critical thickness for the film transition from the NFE state to poly-domain or single-domain FE, FI or AFI states depends on the voltage $U$. Moreover, the critical thickness decreases with the $U$ increase according to our calculations. Thus, the electric field-induced transition to the FI or AFI phases of thin films covered with a layer of ions is possible due to the field stimulation of the electrochemical reaction of ion formation. For instance the 10-nm film revealed the unusual field-induced transition into FI and AFI states induced by the strong field dependence of the ion charge density. Note that electric field induced FE-state has been observed in ferroelectrics since 90ies [47, 48, 49].

The physical origin of the AFI double hysteresis loops characteristic to AFI state can be explained by the dependence of the surface charge density on electric potential at $\Delta G_1^{00} = \Delta G_2^{00} \sim 0.5$ eV, which has the form of Langmuir absorption isotherm [shown in **Fig. 1(b)**]. Since the surface charge is responsible for the polarization screening in a ferroelectric film, high charge densities (corresponding to high acting electric potential $|e\varphi| \gg k_B T$) can provide an effective screening, and the low ones (corresponding to small potentials $|e\varphi| \leq k_B T$) can provide a weak incomplete screening only. At room temperatures the 10-nm PZT film is in the FI state for $0 \leq \Delta G_i^{00} \leq 0.1$ eV, and relatively weak ionic screening of its depolarization field is enough to support the state. Corresponding hysteresis loop has a single ferroelectric-like shape with relatively high remanent polarization and coercive voltage. Polarization behavior changes drastically with $\Delta G_i^{00}$ increase, when the film approaches the AFI or NFE states. A high screening degree leads to higher critical thickness of the transition [**Ошибка! Закладка не определена.**, 1, 15]. Thus, additional screening by ions is urgently required to maintain the thin film in a polarized or antipolarized state. As one can see from **Fig. 1(b)**, the screening increase appears at nonzero potential φ that is in turn proportional to the applied voltage $U$. The critical voltage corresponding to the screening degree enough to suppress the thickness-induced paraelectric transition opens the double hysteresis loops of polarization, which in turn induce the double loops of concentration hysteresis. The situation for thicker films (e.g. 25 nm thick and more) appeared more usual, since the single loop opens at high voltages for all $\Delta G_i^{00}$, and its width and asymmetry much weaker depends on $\Delta G_i^{00}$ values [see **Fig. S3**].

Detailed comparison of **Fig. 4(a)** and **4(b)** brings forth the question of how to explain the strong decrease (up to the absence) of the surface charge at high voltages [(10 – 20)V], if the polarization saturates at the same time [e.g. compare the plots of $P(U)$ and $\sigma(U)$ calculated for $\{\Delta G_1^{00}, \Delta G_2^{00}\} = \{0.1, 0.1\}$]? To disclose the reason we calculate the polarization, surface charge, electric potential and field at the surface of ferroelectric 10-nm PZT film, where the surface ions are located, for a wider range of applied voltages [see **Figs. S4**]. It appeared that the potential becomes very small (and hence the ion charge is small), and the potential and field drop is located in the ultra-thin physical gap of thickness $\lambda$ = 0.4 nm [see **Fig. S4(f)**]. On the other hand, the field in a ferroelectric film is much smaller and display hysteretic behavior [**Fig. S4(e)**]. **Figure S5** is for the same 10-nm PZT film in a deep paraelectric phase (at high temperature).

We have further calculated the voltage dependence of the total charge at bottom ($z = h$) and top ($z = -\lambda$) electrodes, $Q_B(U)$ and $Q_T(U)$, respectively [see **Figs. 5**]. It appeared that the value and hysteresis shape of $Q_B(U)$ almost coincide with the dependence $P(U)$ in the considered quasi-static limit (i.e. at very low frequencies of applied voltage). However $Q_T(U)$ is linear and so it principally differs from $P(U)$ due to the surface charge hysteresis $\sigma(U)$, as well as due to the electric field drop in the ultra-thin dielectric gap that "conduct" only the displacement currents. Counter-clockwise path-tracing of $P(U)$ loops corresponds to clockwise path-tracing of $\sigma(U)$ loops.

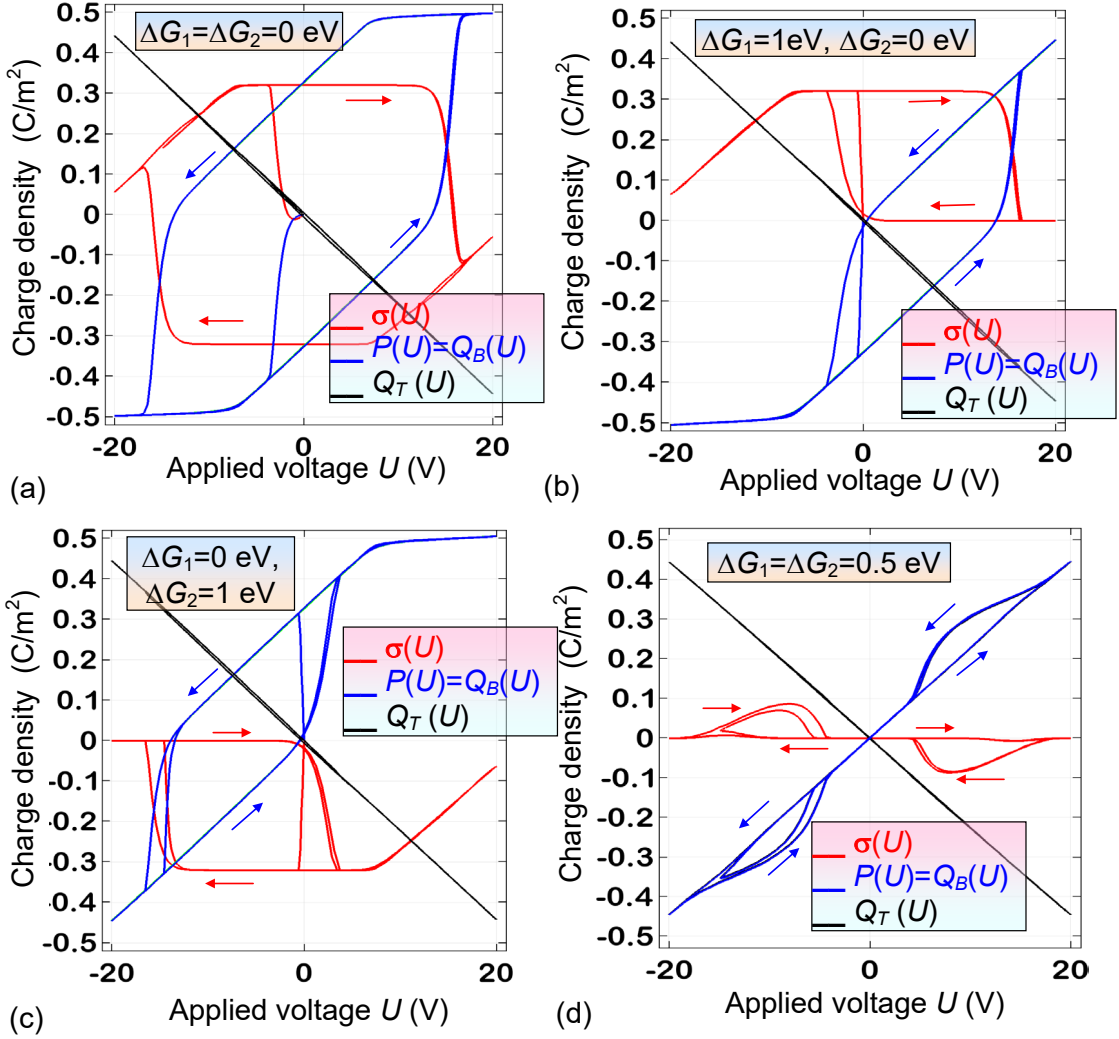

**FIGURE 5.** Dependences of average polarization $P(U)$ and surface charge $\sigma(U)$ on applied voltage U calculated for a 10-nm PZT film with fixed values $\{\Delta G_1^{00}, \Delta G_2^{00}\}$ indicated in the labels at plots **(a)-(d)**. Arrows indicate the direction of the loops path-tracing Temperature T=300 K, other parameters are the same as in **Fig. 2**.

On of the conclusive remarks of the section is that **Figures 2-4** argue that the boundaries between PD-FE, SD-FE, FI and AFI states is not sharp, similarly to the difference between ferroelectric and pyroelectric states. However we can actually separate them based on the thickness evolution of free energy for polarization, as will be demonstrated below.

## IV. FREE ENERGY OF FERRO-IONIC SYSTEM

Here, we develop simplified analytical model to get insight into numerically analyzed behaviors of FI states. Since the stabilization of single-domain polarization in ultrathin perovskite films takes place due to the chemical switching (see e.g. [25, 26, 22-26]), we can consider the ferroelectric film in the SD-FE, FI, AFI, NFE or PE states and assume that polarization distribution $P_3(x,y,z)$ is smooth. In this case, the behavior of the polarization averaged over film thickness $P = \langle P_3 \rangle$ and surface charge density $\sigma$ can be described via the coupled nonlinear algebraic equations derived in Refs. [34 - 36].

Here we consider the stationary case, when $\sigma = \sigma_0$, and also $p_{exc} = p_{atm}$. The free energy $G$ minimization of which gives the coupled equations for polarization dynamics, has the form:

$$\frac{G[P,\Psi]}{S} = h\left(\frac{\alpha_R}{2}P^2 + \frac{\beta}{4}P^4 + \frac{\gamma}{6}P^6\right) - \Psi P - \varepsilon_0\varepsilon_{33}^b \frac{\Psi^2}{2h} - \frac{\varepsilon_0\varepsilon_d}{2}\frac{(\Psi-U)^2}{\lambda} + \int_0^\Psi \sigma_0[\phi]d\phi \quad (3)$$

Coefficient $\alpha_R = \alpha_T(T_C - T) + \frac{g_{33}}{h}\left(\frac{1}{\Lambda_+} + \frac{1}{\Lambda_-}\right)$ is the coefficient $\alpha$ renormalized by "intrinsic" gradient-correlation size effects (the term $\sim g_{33}/h$). The first term in Eq.(3) is Landau-Ginzburg polarization energy. The second term, $\Psi P$, represents the energy of interaction of polarization $P$ with overpotential $\Psi$. The terms $\varepsilon_0\varepsilon_{33}^b\frac{\Psi^2}{2h}$ and $\frac{\varepsilon_0\varepsilon_d}{2}\frac{(\Psi-U)^2}{\lambda}$ are the energies of electric field in the ferroelectric film, and in the gap, correspondingly. The last term, $\int_0^\Psi \sigma_0[\phi]d\phi$, is the surface charge energy. The formal minimization, $\frac{\partial G[P,\Psi]}{\partial P} = 0$ and $\frac{\partial G[P,\Psi]}{\partial \Psi} = 0$, couples the polarization and overpotential as following

$$\alpha_R P + \beta P^3 + \gamma P^5 = \frac{\Psi}{h}, \quad (4a)$$

and

$$\Psi = \frac{\lambda(\sigma - P) + \varepsilon_0\varepsilon_d U}{\varepsilon_0(\varepsilon_d h + \lambda\varepsilon_{33}^b)}h. \quad (4b)$$

The energy given by Eq. (3) has an absolute minimum at high $\Psi$. According to the Biot's variational principle [50], we can further use the incomplete thermodynamic potential, partial minimization of which over $P$ will give the coupled equations of state, and, at the same time, it has an absolute minimum at finite $P$ values. In **Appendix B** we derived the $P$-dependent thermodynamic potential, $F[P]$. Assuming that $|eZ_i\Psi/k_BT| \ll 1$, the series expansion of $F[P]$ on polarization powers has the form:

$$F \approx A_R\frac{P^2}{2} + B_R\frac{P^4}{4} + C_R\frac{P^6}{6} - PE_{eff}. \quad (5a)$$

Here, the renormalized coefficients are

$$A_R = \frac{\lambda}{\varepsilon_0(\varepsilon_d h + \lambda\varepsilon_{33}^b)} + f(\Delta G_i^{00}, h, T)\alpha_R(T), \quad (5b)$$

$$B_R = \beta \cdot f(\Delta G_i^{00}, h, T), \quad C_R = \gamma \cdot f(\Delta G_i^{00}, h, T), \quad (5c)$$

The function $f$ in Eqs. (5b) and (5c) is

$$f\left(\Delta G_i^{00}, h, T\right) = 1 + \frac{\lambda h}{\varepsilon_0\left(\varepsilon_d h + \lambda \varepsilon_{33}^b\right)} \sum_{i=1,2} \frac{(eZ_i)^2}{A_i k_B T} \left(1 + \exp\left(\frac{\Delta G_i^{00}}{k_B T}\right)\right)^{-2}. \tag{6}$$

Because the renormalized coefficients $B_R$ and $C_R$ are always positive, the potential (5) has a local minimum in dependence on $A_R$ sign and effective field strength. The effective field produced by ionic charge has the form:

$$E_{eff}\left(U, \Delta G_i^{00}\right) = \frac{\lambda}{\varepsilon_0\left(\varepsilon_d h + \lambda \varepsilon_{33}^b\right)} \sum_{i=1,2} \frac{eZ_i}{A_i} \left(1 + \exp\left(\frac{\Delta G_i^{00}}{k_B T}\right)\right)^{-1} - \frac{\varepsilon_d U}{\varepsilon_d h + \lambda \varepsilon_{33}^b}, \tag{7}$$

Since usually $\lambda \ll h$ the effective electric field (6) is proportional to $1/h$ and so $E_{eff}$ becomes higher in thinner films. To study the polarization reversal $E_{eff}$ should be compared with the intrinsic thermodynamic coercive field $E_c$ for the actual range of temperatures and thickness. To determine $E_c$, we solve elementary equation of state, $A_R P + B_R P^3 + C_R P^5 = E_{eff}$. The field $E_C$ defined directly from the equation has the form:

$$E_C = \frac{2}{5}\left(2B_R + \sqrt{9B_R^2 - 20 A_R C_R}\right)\left(\frac{2A_R}{-3B_R - \sqrt{9B_R^2 - 20 A_R C_R}}\right)^{3/2}. \tag{8}$$

The coercive field $E_C = \frac{2}{3\sqrt{3}}\sqrt{-\frac{A_R^3}{B_R}}$ at $C_R = 0$. Because the coefficients $A_R$, $B_R$ and $C_R$ are thickness-dependent, the coercive field is thickness-dependent as well.

We further analyze the dependences of the effective field, free energy and polarization states on the ion formation energies, temperature and film thickness. The effective field $E_{eff}$ induced by surface ions was calculated at $U=0$ as a function of the surface ions formation energies $\Delta G_i^{00}$ for PZT film thickness $h$ = 10, 50 and 100 nm, as shown in **Fig. 3(d)**. The contours of constant field delineate the boundary between FI and SD-FE states in 100-nm and 50-nm thick PZT films [see dashed curves in **Figs. 3(a)** and **3(b)**], as well as the field defines the boundary between FI and NFE states in thin 10-nm film [shown by dashed curves in **Fig. 3(c)**], [51]. This confirms the conjecture that the fields induced by surface ions can polarize thin ferroelectric films and prevent the domain formation for the films of thickness less than critical.

Note that the effective field [see Eq. (7)] is anti-symmetric with respect to $\Delta G$, namely $E_{eff}(0, \Delta G) = -E_{eff}(\Delta G, 0)$. The dependence of $E_{eff}\left(\Delta G_1^{00}, \Delta G_2^{00}\right)$ vs. $\Delta G_2^{00}$ was calculated from Eq. (7) for $\Delta G_1^{00} = 0$, $U=0$ and different thickness $h$ = (10 – 100) nm PZT film [see different curves in **Fig. 6(a)**]. The field rapidly increases with $\Delta G_2^{00}$ increase from 0 to 0.1 eV and then saturates. At that the saturation $E_{eff}$ values increase with the film thickness decrease from 0.08 V/nm for 100-nm film to 0.5 V/nm for 10-nm film. The thickness dependence of $E_{eff}(\Delta G_1^{00}, \Delta G_2^{00})$ calculated for several $\{\Delta G_1^{00}, \Delta G_2^{00}\}$ in comparison with the thermodynamic coercive fields $E_C[h]$ is shown in **Fig. 6(b)**. It is seen that the the coercive field abruptly vanishes with the film thickness decrease, and the critical thickness for which this happens depends on the values of $\{\Delta G_1^{00}, \Delta G_2^{00}\}$; in particular it increases with the $\Delta G_1^{00}$ increase at fixed $\Delta G_2^{00} = 1\,\text{eV}$ [see also **Figs. S6**]. The effective field exists in ultrathin film, and here it formally dominates over zero coercive field.

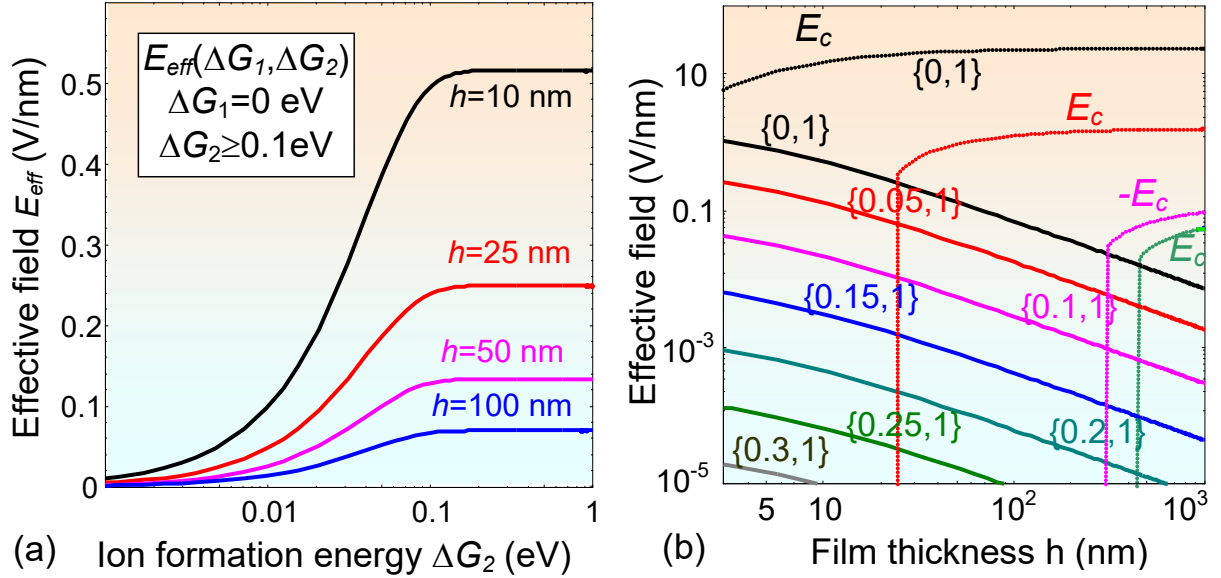

**FIGURE 6. (a)** Effective field $E_{eff}(\Delta G_1^{00}, \Delta G_2^{00})$ depending on the surface ions formation energy $\Delta G_2^{00}$ calculated at $\Delta G_1^{00} = 0$. Different curves correspond to PZT film thickness $h$ = 10, 25, 50 and 100 nm (as indicated by legends). **(b)** Thickness dependence of effective field $E_{eff}(\Delta G_1^{00}, \Delta G_2^{00})$ calculated for several pairs of values $\{\Delta G_1^{00}, \Delta G_2^{00}\}$ (legends near the curves in eV, where $\Delta G_1^{00}$ varies and $\Delta G_2^{00} = 1\,\text{eV}$). Coercive fields $E_C[h]$ are shown by dashed curves. Temperature $T$ = 300 K, $U$ = 0 and other parameters are the same as in **Fig. 3.**

Elementary analyses of the algebraic equation $A_R P + B_R P^3 + C_R P^5 = E_{eff}$, the mathematical form of which coincides with a conventional Landau equation in external field $E_0$, shows that in the case of the purely second order phase transitions ( $A_R < 0$, $B_R > 0$, $C_R = 0$ ) the equation has three real roots. The one of these roots is absolutely unstable, and two others are absolutely stable or metastable in dependence on the temperature, applied voltage and film thickness [see global and local minima and maxima in **Fig. 7(a)** and **7(b)** for 100-nm and 50-nm films at room temperature]. In particular, the only one real root, $P \approx E_{eff}/A_R$, exists at $A_R > 0$ for small thicknesses and/or high temperatures [see global minimum in **Fig. 7(b)** for 10-nm films at room temperature]. Dependence of the absolutely stable polar state $P_1$ on the surface ions formation energies $\Delta G_i^{00}$ calculated for the film thickness $h = 100$, 50 and 10 nm are shown in **Figs. 7(c),(d)** and **(e)** respectively. Notice, that the polar states $P_i$ are significantly $\Delta G_i^{00}$-dependent only at $0 \leq \Delta G_i^{00} \leq 0.1\,\text{eV}$. Stable, metastable and unstable polar states in dependence on the surface ions formation energies $\Delta G_i^{00}$ calculated for PZT film thickness $h = 10$, 50, 100 and 300 nm and $T = 600$, 300 K are shown in **Figs. S7-S9**.

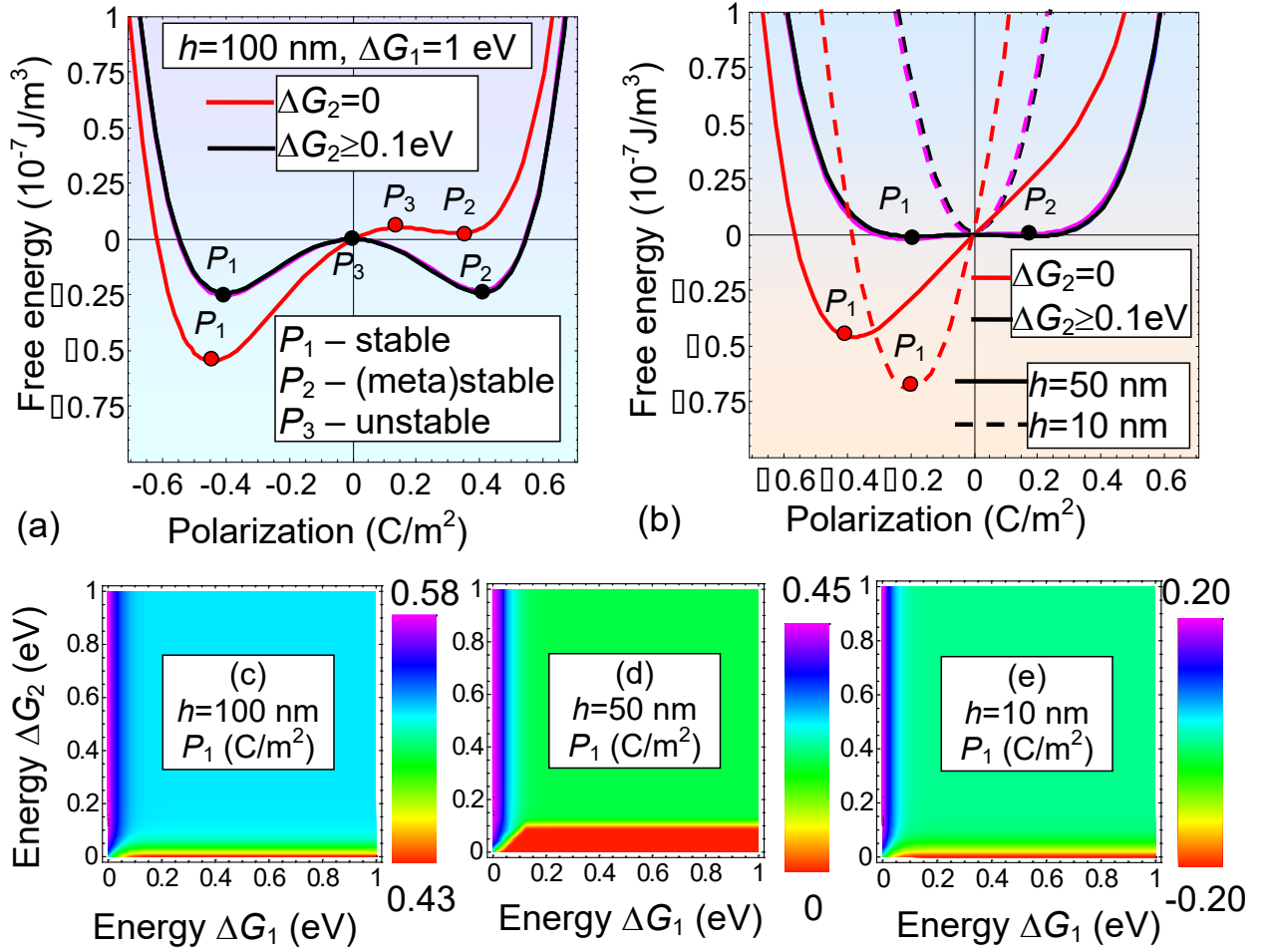

**FIGURE 7.** Dependences of the free energy on average polarization calculated in the case of screening by ions, zero applied voltage and different values of PZT film thickness $h = 100$ nm **(a)**, 50 (solid curves) and 10 (dashed curves) nm **(b)**. Each curve corresponds to $\Delta G_1^{00} = 1\,\text{eV}$ and $\Delta G_2^{00} = 0$ (red) or $\Delta G_2^{00} = (0.1 - 1)\,\text{eV}$ (black an other colors, which coincide). **(c)-(e)** Dependence of the absolutely stable polar state $P_1$ on the surface ions formation energies $\Delta G_i^{00}$ calculated for $h = 100$, 50 and 10 nm (legends at the plots). Color scale bar is polarization in C/m². Temperature $T = 300$ K, $U = 0$ and other parameters are the same as in **Fig. 3**.

Because the coefficients $A_R$, $B_R$ and $C_R$ depend on the ion formation energies $\Delta G_i^{00}$ as per Eqs. (5), the free energy (5) is $\Delta G_i^{00}$-dependent, and **Figs. 8** demonstrate how strong and nontrivial the dependence is for different film thicknesses. For 300-nm film, the two deep absolute and local minima, and one high maximum (barrier), which depth and position are $\Delta G_2^{00}$-dependent at $0 \leq \Delta G_2^{00} \leq 0.1 \text{eV}$ and fixed $\Delta G_1^{00} = 1\text{eV}$ [**Fig. 8(a)**]. The minima become much shallower and barrier almost disappear with the film thickness decrease up to 50 nm [compare the maps in **Fig. 8(b)** and **8(c)**]. Only one absolute minima exists for 10-nm film [**Fig. 8(d)**].

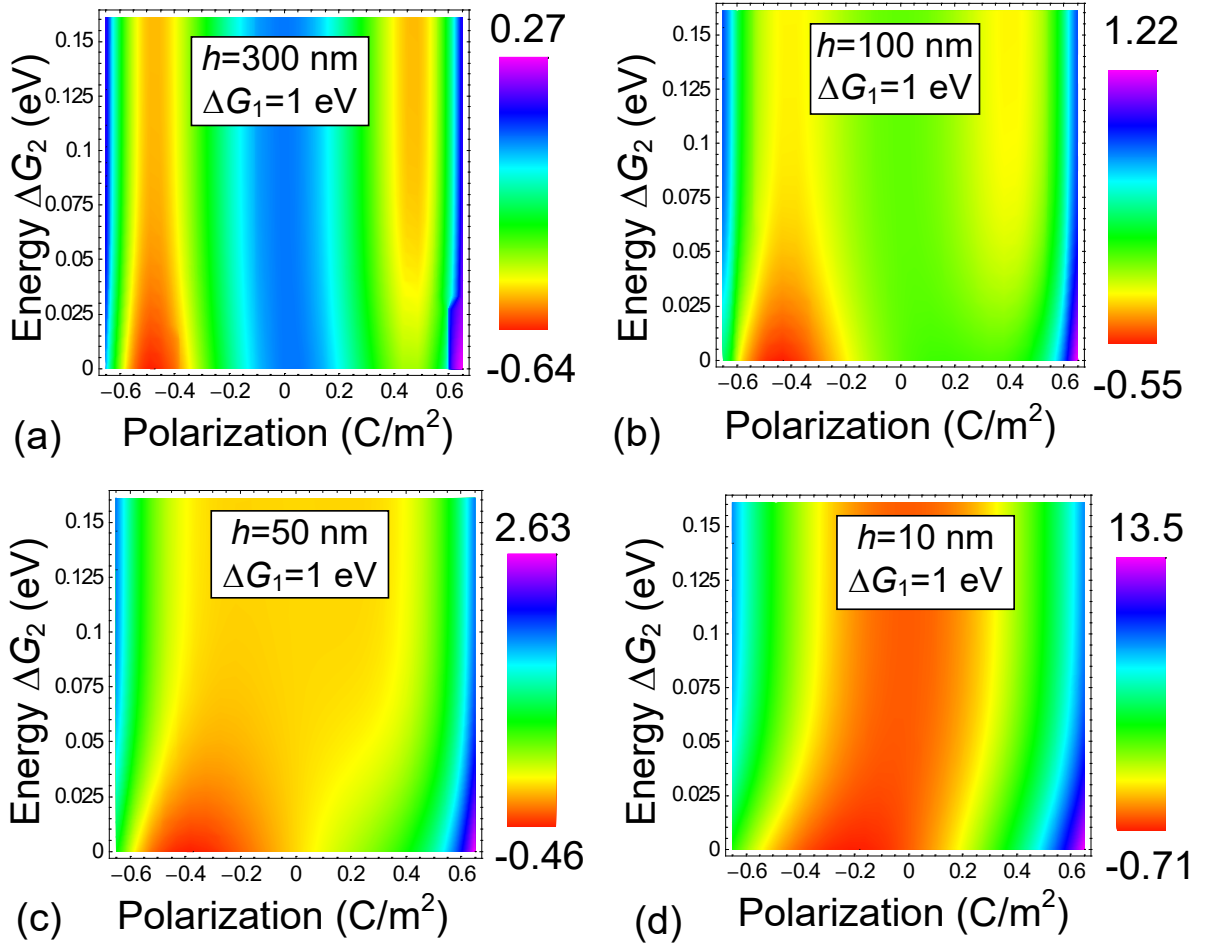

**FIGURE 8.** Dependence of the free on the average polarization and surface ions formation energy $\Delta G_2^{00}$ calculated for $\Delta G_1^{00} = 1\text{eV}$, PZT film thickness $h$ = 300 nm **(a)**, 100 nm **(b)**, 50 nm **(c)** and 10 nm **(d)**. Color scale bar is the energy in $10^7$ J/m$^3$. Temperature $T$ = 300 K, $U$ = 0 and other parameters are the same as in **Fig.3**.

**Figure 9** illustrates the dependence of the free energy on the average polarization $P$ and film thickness $h$ calculated for $\Delta G_2^{00} = 0$ [**Fig. 9(a)**] and $\Delta G_2^{00} = 0.1\text{eV}$ [**Fig. 9(a)**] at fixed $\Delta G_1^{00} = 1\text{eV}$ and room temperature. Two nonequivalent minima of polarization ($P_1$ and $P_2$) are separated by a barrier ($P_3$ state) at $\Delta G_2^{00} = 0$ and fixed $h$ above 100 nm, and the only minimum remains for thin films of e.g. 10 nm thick [see polarization points in **Fig. 9(a)**]. Two almost equivalent minima of polarization ($P_1$ and $P_2$) are separated by a barrier ($P_3$ state) at $\Delta G_2^{00} = 0.1\text{eV}$ and fixed $h$ more than 50 nm, and they transforms into a saddle point with the film thickness decrease [see polarization points in **Fig. 9(b)**]. In fact, **Figs. 9** argues that the difference between SD-FE and FI states is not sharp, similarly to the difference between ferroelectric and pyroelectric states, but we can actually separate them based on the thickness evolution of free energy for polarization. For thick films there are 2 minima, and they are almost the same as in the bulk. For small thickness we start to see the mergence of the minima and disappearance of corresponding polarization state and so the ferro-ionic system undergoes second order transition with the thickness decrease.

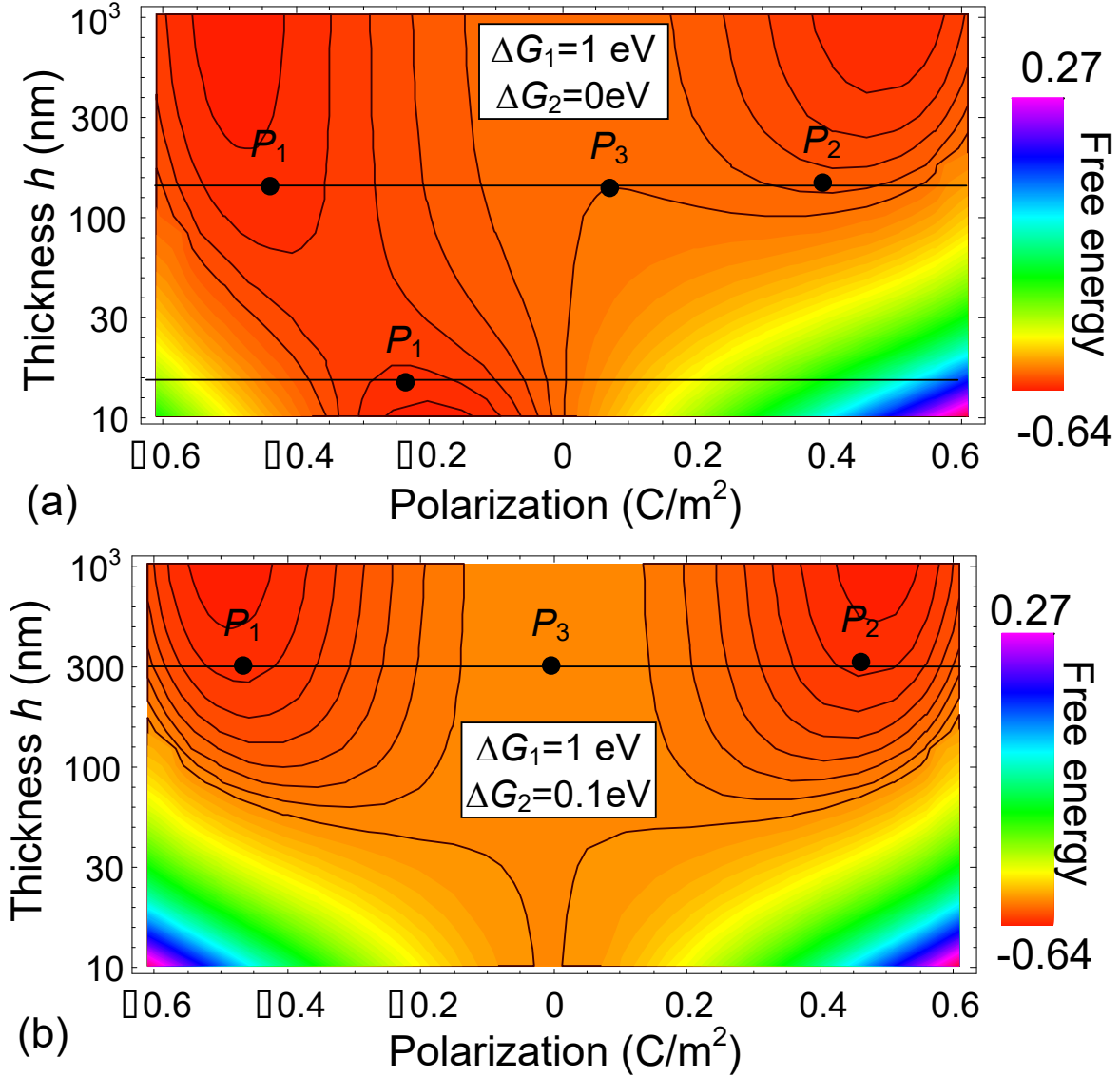

**FIGURE 9.** Dependence of the free energy on the average polarization and PZT film thickness calculated for $\Delta G_1^{00} = 1\,\text{eV}$, $\Delta G_2^{00} = 0$ **(a)** and $\Delta G_2^{00} = 0.1\,\text{eV}$ **(b)**. Color scale bar is the energy in $10^7\,\text{J/m}^3$. Temperature $T = 300$ K, $U = 0$ and other parameters are the same as in **Fig. 3.**

The temperature and $\Delta G_2^{00}$ dependences of three polarization states for 100-nm film, corresponding to the roots of algebraic equation $A_R P + B_R P^3 + C_R P^5 = E_{eff}$ for the case of the purely second order phase transitions ($A_R < 0$, $B_R > 0$, $C_R = 0$), are shown by red, magenta and blue curves in **Fig. 10(a)**. Each curve in the group (denoted by single color) corresponds to the definite value of $\Delta G_2^{00}$ within the range (0 - 0.9) eV. The only one stable polar state, $P \approx E_{eff}/A_R$ that exists in a 10-nm film [because $A_R > 0$ for the case] is shown **Fig. 10(b)**. Corresponding dependences of the polarization states on the surface ions formation energy $\Delta G_2^{00}$ calculated for $T$ = 300 K and 600 K are shown in **Figs. 10(c)** and **10(d),** respectively.

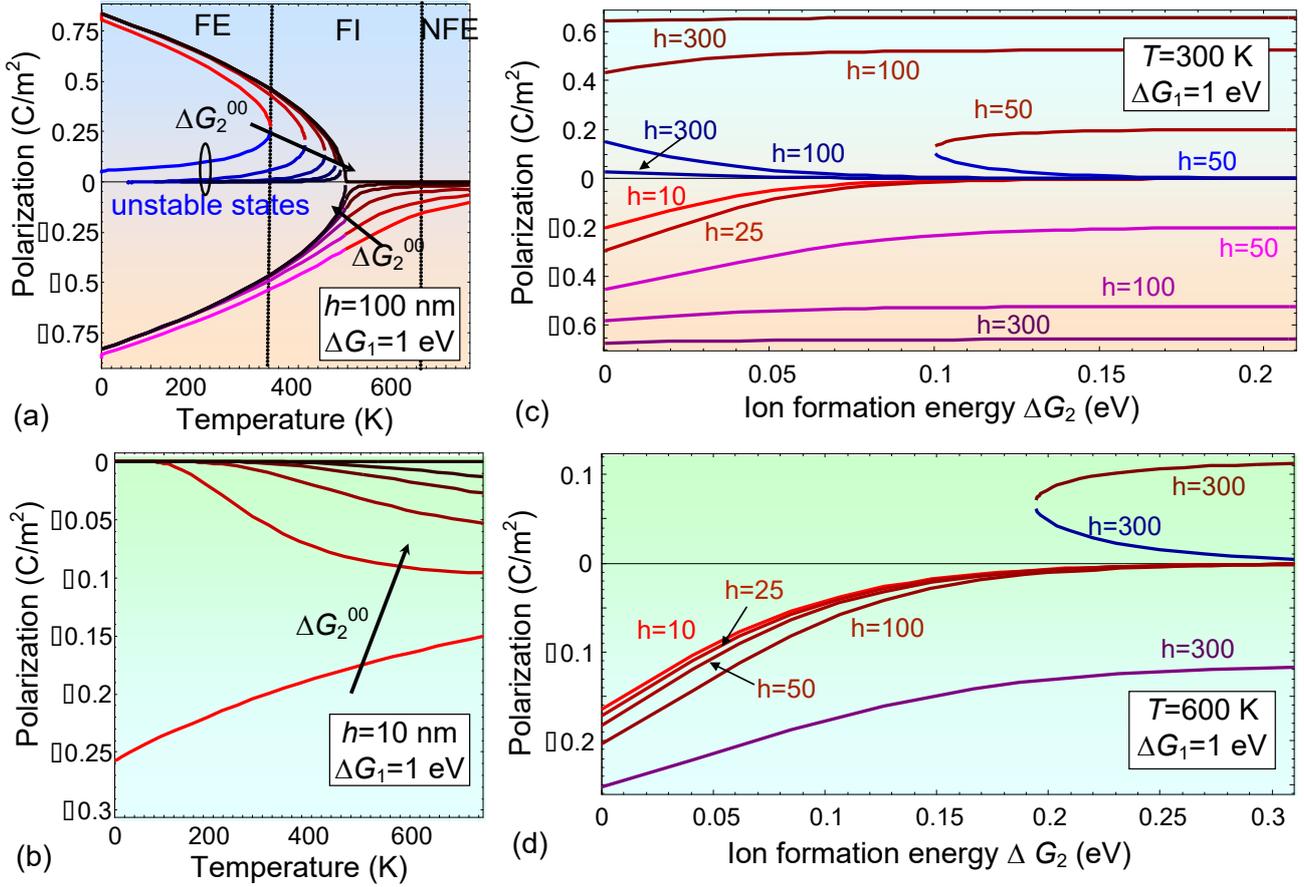

**FIGURE 10. (a-b)** Temperature dependences of the average polarization calculated in the case of screening by ions**,** zero applied voltage and different values of film thickness $h$ = 100 nm **(a)** and 10 nm **(b)**. Each curve in the one-colored group corresponds to the definite value of $\Delta G_2^{00}$ = (0, 0.05, 0.1, 0.15, 0.2, 0.9) eV. The arrows indicate the direction of $\Delta G_2^{00}$ increase. Unstable branches (inside the ellipse) are shown by blue colors. **(c-d)** Dependence of the polarization states on the surface ions formation energy $\Delta G_2^{00}$ calculated for $T$ = 300 K **(c)** and 600 K **(d).** Different curves correspond to PZT film thickness $h$ = 10, 25, 50, 100 and 300 nm (as indicated by legends in nm). Stable polar states [P$_1$ and P$_2$ in **Fig. 6(a,b)**] are shown by red and magenta colors. Unstable states [P$_3$ in **Fig. 6(a,b)**] existing for $h \geq 50$ nm; they are shown by blue colors. The energy $\Delta G_1^{00}$ = 1eV, $U$ = 0 and other parameters are the same as in **Fig. 3.**

# IV. DISCUSSION AND OUTLOOK

We established the role of the surface ions formation energies on the polarization states and its reversal scenarios, domain structure and corresponding phase diagrams of ferroelectric thin films. At that it appeared that the role of the physical gaps on the domain formation and stability is critical. Using 3D finite elements modeling we analyze the distribution and hysteresis loops of ferroelectric polarization and ionic charge, and dynamics of the domain states. Obtained results delineate the regions of single- and poly-domain ferroelectric, ferroionic, antiferroionic and non-ferroelectric states as a function of surface ions formation energies $\Delta G_i^{00}$, film thickness $h$, applied voltage $U$ and temperature $T$.

We revealed the unusual dependence of the film polar state and domain structure parameters on the ion formation energy value and, that is more unexpected, on the applied voltage. In particular we observed the voltage-induced phase transitions into the single-domain ferroionic or antiferroionic states in a thin film covered with ion layer of electrochemically active nature, when the energies $\Delta G_i^{00}$ are significantly different. Ferroionic poly-domain states have not been realized in ultrathin films. At that the physical origin of the double "antiferroionc" hysteresis loops of polarization and surface charge revealed at close ions formation energies ($\left|\Delta G_1^{00} - \Delta G_2^{00}\right| \leq 0.05\,\text{eV}$) can be explained by the dependence of the surface charge density on electric potential, which has the form of Langmuir absorption isotherm.

We further map the analytical theory for 1D system onto an effective Landau-Ginzburg free energy and establish the correspondence between the 3D numerical and 1D analytical results at zero applied voltage. The results argues that the difference between the single domain ferroelectric and different ferroionic states is not sharp, similarly to the difference between ferroelectric and pyroelectric states, and we can actually separate them based on the thickness dependence of ferro-ionic system free energy. For thick films ($h \geq 100$ nm) there are 2 minima, and they are almost the same as in the bulk. For small thickness ($h < 50$ nm) we start to see the mergence of the minima and disappearance of corresponding polarization state. In result the ferro-ionic system undergoes the sort of the second order transition taking place with the thickness $h$ decrease.

This approach allows getting the overview of the phase diagrams of ferro-ionic systems, and exploring the specific features of polarization reversal and domain evolution phenomena. On the other hand many important questions remain for further studies.

At first, if the concentration of the ions becomes smaller, the distance between them increases as its square root. When the distance between the surface ions becomes compatible (or even smaller) than the film thickness, each of the ions should be considered as a point charged defect and the continuous approximation for their charge density $\sigma_0[\varphi]$ becomes invalid. Complementary atomistic calculations should be performed n the case of small ion concentration and/or high saturation densities of the surface ions defining their steric limit (saturation area per one ion).

Secondly a detailed semi-phenomenological and semi-microscopic study of the influence of the electrochemical charges parameters, such as their type (oxygen anions and vacancies vs. other charges such as protons, hydroxyls, etc), enthalpy and/or formation energies seem to be urgent. Notice that the study can reveal that the steric limit value and all other properties related with their thermal activation (analog of the and $\alpha(\varphi)$) including the net form of the potential dependence of the occupied cites (analog of the and $\theta_i(\varphi)$) that in principle can be different from Langmuir isotherm (e.g. Frumkin isothermal function [52]).

In the third, the dynamical processes, including the polarization and electromechanical response relaxation and their periodic change (cross-over between hysteretic and non-hysteretic behaviors) under the periodic applied voltage are poorly studied. An expected results analysis is impossible without an establishment of the correct hierarchy of different charge species relaxation times.

**Acknowledgements.** The publication contains the results of studies conducted by President's of Ukraine grant for competitive projects (grant number F74/25879) of the State Fund for Fundamental Research (A.N.M., E.A.E., A.I.K.). S.V.K. acknowledges the Office of Basic Energy Sciences, U.S. Department of Energy. Part of work was performed at the Center for Nanophase Materials Sciences, which is a DOE Office of Science User Facility.

# APPENDIX A. ELECTROSTATIC EQUATIONS WITH BOUNDARY CONDITIONS

Quasi-static electric field inside the ferroelectric film is defined via electric potential as $E_3 = -\partial \varphi_f / \partial x_3$, where the potential $\varphi_f$ satisfies electrostatic equations for each of the medium (gap and ferroelectric film) acquires the form:

$$\Delta \varphi_d = 0, \qquad \text{(inside the gap } -\lambda \leq z \leq 0\text{)} \tag{A.1a}$$

$$\left( \varepsilon_{33}^b \frac{\partial^2}{\partial z^2} + \varepsilon_{11}^f \Delta_\perp \right) \varphi_f = \frac{1}{\varepsilon_0} \frac{\partial P_3^f}{\partial z}, \qquad \text{(inside the ferroelectric film } 0 < z < h\text{)} \tag{A.1b}$$

where $\Delta$ is 3D-Laplace operator is $\Delta$, $\Delta_\perp$ is 2D-Laplace operator.

Boundary conditions (**BCs**) to the system (A.1) have the form:

$$\varphi_d\big|_{z=-\lambda} = U, \quad \left( \varphi_d - \varphi_f \right)\big|_{z=0} = 0, \quad \varphi_f\big|_{z=h} = 0, \tag{A.2a}$$

$$\left( \varepsilon_0 \varepsilon_d \frac{\partial \varphi_d}{\partial z} + P_3^f - \varepsilon_0 \varepsilon_{33}^b \frac{\partial \varphi_f}{\partial z} - \sigma \right)\bigg|_{z=0} = 0. \tag{A.2b}$$

# APPENDIX B. FREE ENERGY WITH RENORMALIZED COEFFICIENTS

If we assume that polarization distribution $P_3(x,y,z)$ is smooth enough, the coupled nonlinear algebraic equations for the polarization averaged over film thickness $P = \langle P_3 \rangle$ and surface charge density $\sigma$ are valid [34-37]:

$$\Gamma \frac{\partial P}{\partial t} + \alpha_R P + \beta P^3 + \gamma P^5 = \frac{\Psi(U,\sigma,P)}{h}, \quad \tau \frac{\partial \sigma}{\partial t} + \sigma = \sigma_0[\Psi(U,\sigma,P)]. \tag{B.1}$$

The overpotential is given by expression $\Psi(U,\sigma,P) = \dfrac{\lambda(\sigma - P) + \varepsilon_0 \varepsilon_d U}{\varepsilon_0 \left( \varepsilon_d h + \lambda \varepsilon_{33}^b \right)} h$ and the function

$\sigma_0[\psi] = \sum_i \dfrac{eZ_i \theta_i(\psi)}{A_i} \equiv \sum_i \dfrac{eZ_i}{A_i} \left( 1 + \exp\left( \dfrac{\Delta G_i^{00} + eZ_i \psi}{k_B T} \right) \right)^{-1}$. Electric potentials acting in the dielectric gap ($\varphi_d$) and in the ferroelectric film ($\varphi_f$) linearly depends on the coordinate z and overpotential, namely $\varphi_d = U - \dfrac{z+\lambda}{\lambda}(U - \Psi)$ and $\varphi_f = (h-z)\dfrac{\Psi}{h}$ (see **Fig. 1**).

Next we consider the stationary case when one can put $\sigma = \sigma_0[\Psi(U,\sigma,P)]$ in Eqs.(B.1) and (3). Corresponding free energy $G[P,\Psi]$, which formal minimization, $\dfrac{\partial G[P,\Psi]}{\partial P} = 0$ and $\dfrac{\partial G[P,\Psi]}{\partial \Psi} = 0$, leads to Eqs.(B.1), has the form:

$$\frac{G[P,\Psi]}{S} = h\left( \frac{\alpha_R}{2} P^2 + \frac{\beta}{4} P^4 + \frac{\gamma}{6} P^6 \right) - \Psi P - \varepsilon_0 \varepsilon_{33}^b \frac{\Psi^2}{2h} - \frac{\varepsilon_0 \varepsilon_d}{2} \frac{(\Psi - U)^2}{\lambda} + \int_0^\Psi \sigma_0[\phi] d\phi \tag{B.2}$$

The energy (B.2) has absolute minima at high $\Psi$ values. So, according to the Biot's variational principle, let us find for the incomplete thermodynamic potential, which partial minimization over $P$ will give the equations of state, and, at the same time, it has an absolute minimum at finite $P$ values. For the purpose the first of Eq.(B.1) can be considered as an expression for the overpotential dependence on $P$, i.e., $\Psi[P] = h(\alpha_R P + \beta P^3 + \gamma P^5)$. Substituting here and the above expression for the overpotential $\dfrac{\Psi}{h} = \dfrac{\lambda(\sigma_0[\Psi] - P) + \varepsilon_0 \varepsilon_d U}{\varepsilon_0(\varepsilon_d h + \lambda \varepsilon_{33}^b)}$ we derived the single equation for the average polarization:

$$\alpha_R P + \beta P^3 + \gamma P^5 = \frac{\lambda}{\varepsilon_0(\varepsilon_d h + \lambda \varepsilon_{33}^b)} \sum_{i=1,2} \frac{eZ_i}{A_i} \left(1 + \exp\left[\frac{\Delta G_i^{00} + eZ_i h(\alpha_R P + \beta P^3 + \gamma P^5)}{k_B T}\right]\right)^{-1} - \frac{\lambda P - \varepsilon_0 \varepsilon_d U}{\varepsilon_0(\varepsilon_d h + \lambda \varepsilon_{33}^b)}$$

(B.3)

Corresponding potential that minimization over $P$ gives Eq.(9b) has the form:

$$F[P] = \left(\left(\frac{\lambda}{\varepsilon_0(\varepsilon_d h + \lambda \varepsilon_{33}^b)} + \alpha_R\right)\frac{P^2}{2} + \beta\frac{P^4}{4} + \gamma\frac{P^6}{6} + \frac{\varepsilon_d U}{\varepsilon_d h + \lambda \varepsilon_{33}^b} P \right. \\ \left. - \frac{\lambda}{\varepsilon_0(\varepsilon_d h + \lambda \varepsilon_{33}^b)} \sum_{i=1,2} \frac{eZ_i}{A_i} \int_0^P dp \left(1 + \exp\left(\frac{\Delta G_i^{00} + eZ_i h(\alpha_R p + \beta p^3 + \gamma p^5)}{k_B T}\right)\right)^{-1}\right)$$

(B.4)

Assuming that $\left(1 + \exp\left(\dfrac{\Delta G_i^{00} + eZ_i \Psi}{k_B T}\right)\right)^{-1} \approx \left(1 + \exp\left(\dfrac{\Delta G_i^{00}}{k_B T}\right)\right)^{-1} \left(1 - \dfrac{eZ_i \Psi}{k_B T(1 + \exp(\Delta G_i^{00}/k_B T))}\right)$ under the condition $\left|\dfrac{eZ_i \Psi}{k_B T}\right| \ll 1$ we derived Eqs.(5) - (7).

# Supplementary Materials to

# Effect of surface ionic screening on polarization reversal scenario in ferroelectric thin films: crossover from ferroionic to antiferroionic states

Anna N. Morozovska[1,2]*, Eugene A. Eliseev[3], Anatolii I. Kurchak[4], Nicholas V. Morozovsky[1], Rama K. Vasudevan[5], Maksym V. Strikha[4,6], and Sergei V. Kalinin[5]†

[1] *Institute of Physics, National Academy of Sciences of Ukraine,*
*pr. Nauky 46, 03028 Kyiv, Ukraine*

[2]*Bogolyubov Institute for Theoretical Physics, National Academy of Sciences of Ukraine,*
*14-b Metrolohichna str. 03680 Kyiv, Ukraine*

[3]*Institute for Problems of Materials Science, National Academy of Sciences of Ukraine, Krjijanovskogo 3, 03142 Kyiv, Ukraine*

[4] *V.Lashkariov Institute of Semiconductor Physics, National Academy of Sciences of Ukraine,*
*pr. Nauky 41, 03028 Kyiv, Ukraine*

[5] *The Center for Nanophase Materials Sciences, Oak Ridge National Laboratory,*
*Oak Ridge, TN 37831*

[6] *Taras Shevchenko Kyiv National University, Radiophysical Faculty*
*pr. Akademika Hlushkova 4g, 03022 Kyiv, Ukraine*

---

* corresponding author, e-mail: anna.n.morozovska@gmail.com

† corresponding author2, e-mail: sergei2@ornl.gov

**TABLE SI. Description of physical variables and their numerical values**

| Description of main physical quantities used in Eqs.(1)-(3) | Designation and dimensionality | Value for a structure BaTiO$_3$ film / ionic charge / gap / tip |
|---|---|---|
| Polarization of ferroelectric along polar axis Z | $P_3$ (C/m$^2$) | variable (0.26 for a bulk material) |
| Electric field | $E_3$ (V/m) | variable |
| Electrostatic potentials of dielectric gap and ferroelectric film | $\varphi_d$ and $\varphi_f$ (V) | variables |
| Electric voltage on the tip | $U$ (V) | variable |
| Coefficient of LGD functional | $a_3 = \alpha_T(T - T_C)$ (C$^{-2}$ J m) | T-dependent variable |
| Dielectric stiffness | $\alpha_T$ ($\times 10^5$ C$^{-2}$·J·m/K) | 6.68 |
| Curie temperature | $T_C$ (K) | 381 |
| Coefficient of LGD functional | $\beta$ ($\times 10^9$ J C$^{-4}$·m$^5$) | $-8.18 + 0.01876 \times T$ |
| Coefficient of LGD functional | $\gamma$ ($\times 10^{11}$ J C$^{-6}$·m$^9$) | $1.467 - 0.00331 T$ |
| Gradient coefficient | $g$ ($\times 10^{-10}$ m/F) | (0.5-5) |
| Kinetic coefficient | $\Gamma$ (s× C$^{-2}$ J m) | rather small |
| Landau-Khalatnikov relaxation time | $\tau_K$ (s) | $10^{-11} - 10^{-13}$ (far from Tc) |
| Thickness of ferroelectric layer | $h$ (nm) | variable 3 – 500 |
| Background permittivity of ferroelectric | $\varepsilon_{33}^b$ (dimensionless) | 10 |
| Extrapolation lengths | $\Lambda_-$, $\Lambda_+$ (angstroms) | $\Lambda_- = 1$ Å, $\Lambda_+ = 2$ Å |
| Surface charge density | $\sigma(\varphi, t)$ (C/m$^2$) | variable |
| Equilibrium surface charge density | $\sigma_0(\varphi)$ (C/m$^2$) | variable |
| Occupation degree of surface ions | $\theta_i$ (dimensionless) | variable |
| Oxygen partial pressure | $p_{O2}$ (bar) | 1 (atmospheric) |
| Surface charge relaxation time | $\tau$ (s) | >> Landau-Khalatnikov time |
| Thickness of dielectric gap | $\lambda$ (nm) | 0.4 |
| Permittivity of the dielectric gap | $\varepsilon_d$ (dimensionless) | 1 - 10 |
| Universal dielectric constant | $\varepsilon_0$ (F/m) | $8.85 \times 10^{-12}$ |
| Electron charge | $e$ (C) | $1.6 \times 10^{-19}$ |
| Ionization degree of the surface ions | $Z_i$ (dimensionless) | $Z_1 = +2$, $Z_2 = -2$ |
| Number of surface ions created per oxygen molecule | $n_i$ (dimensionless) | $n_1 = 2$, $n_2 = -2$ |
| Saturation area of the surface ions | $A_i$ (m$^2$) | $A_1 = A_2 = 10^{-18} - 10^{-19}$ |
| Surface defect/ion formation energy | $\Delta G_i^{00}$ (eV) | 0 - 1 |

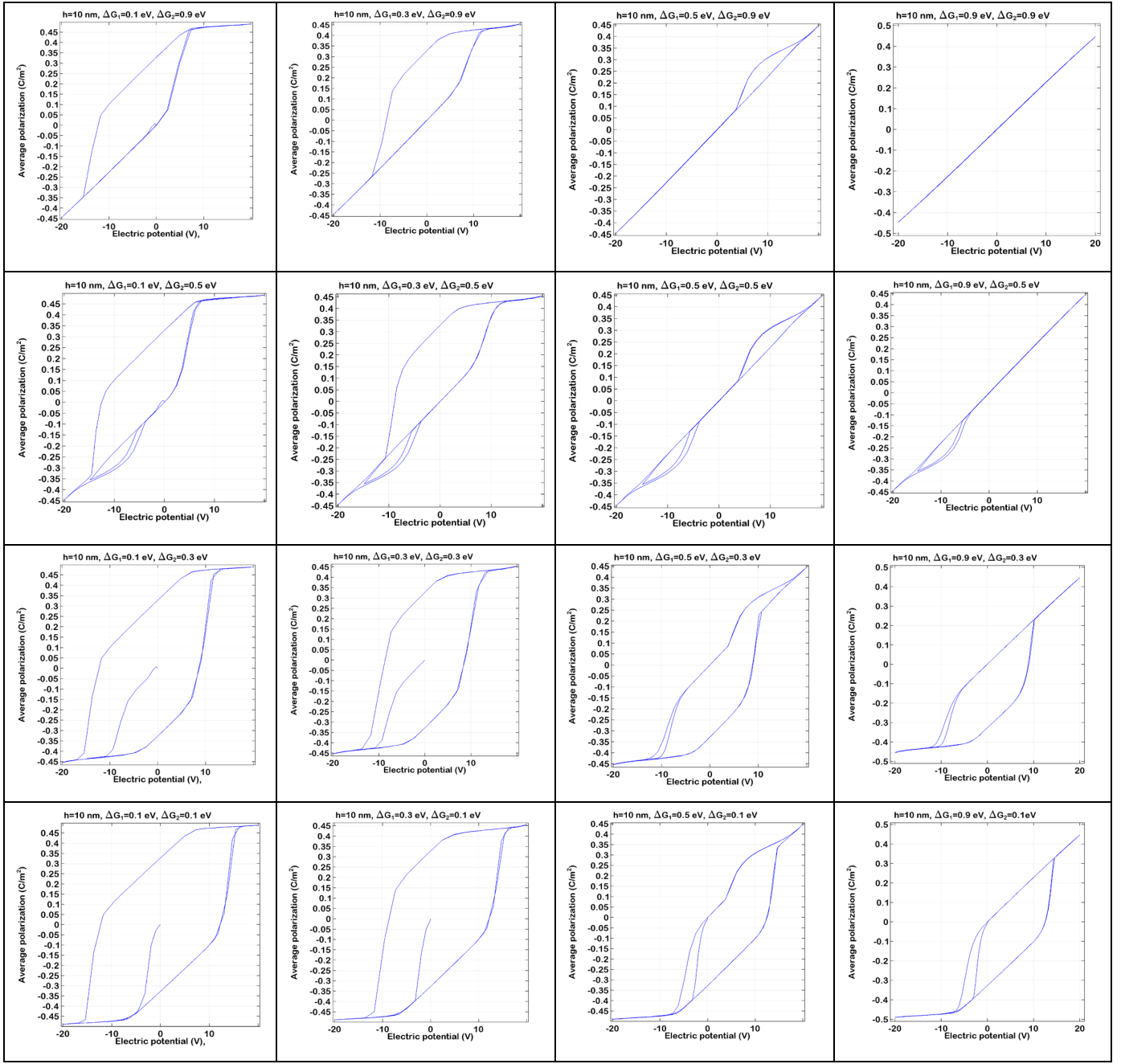

**FIGURE S1.** Dependences $P(U)$ calculated at fixed $\{\Delta G_1^{00}, \Delta G_2^{00}\}$ indicated in the labels for a 10-nm PZT film. Permittivity of the dielectric gap is $\varepsilon_d$=1, its thickness λ=0.4 nm and saturation area of the surface ions $A_1 = A_2 = 10^{-18}$ m², $T$= 300 K, other parameters are listed in **Table SI.**

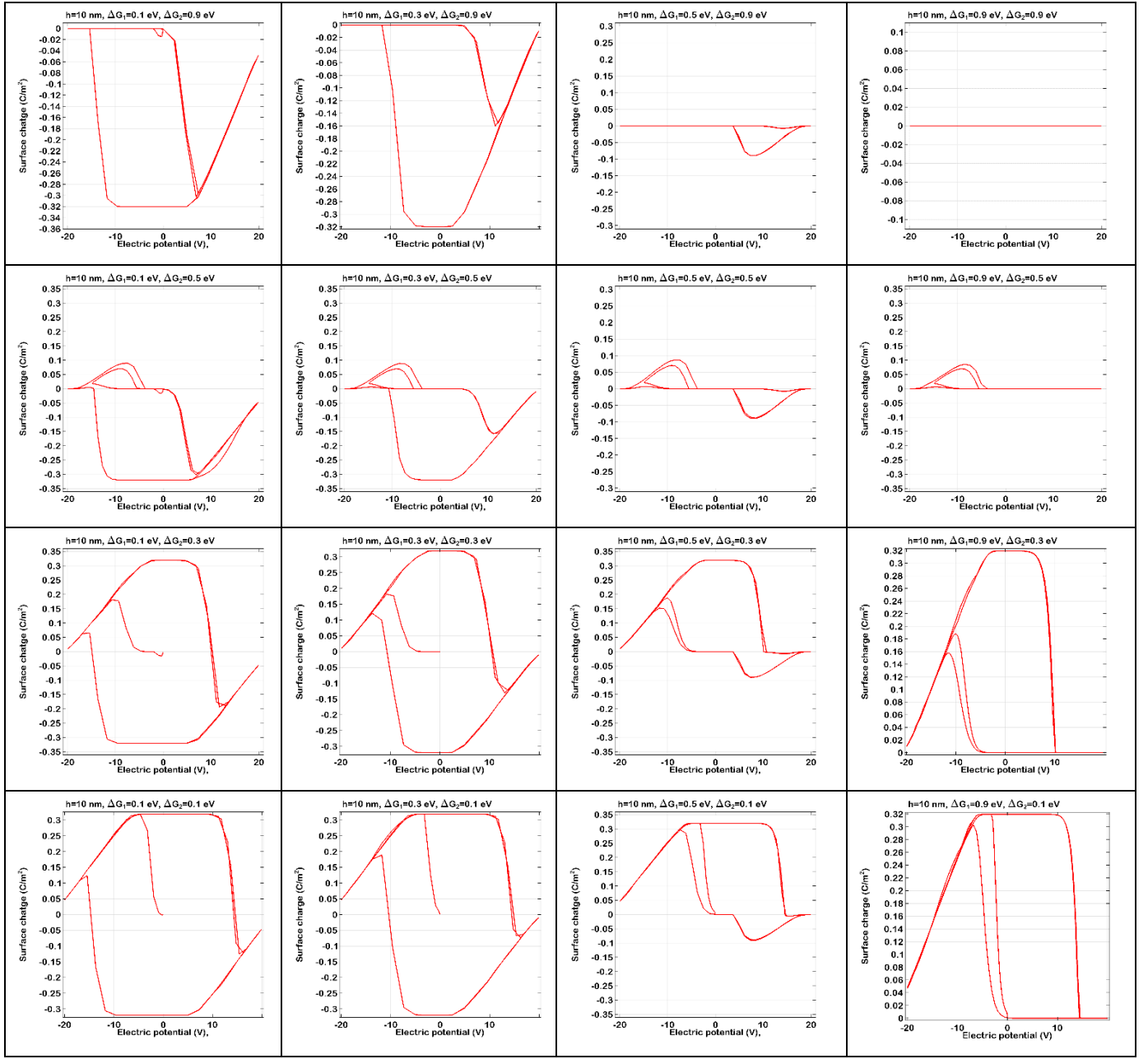

**FIGURE S2.** Dependences $\sigma(U)$ calculated at fixed $\{\Delta G_1^{00}, \Delta G_2^{00}\}$ indicated in the labels for a 10-nm PZT film. Permittivity of the dielectric gap is $\varepsilon_d =1$, its thickness $\lambda=0.4$ nm and saturation area of the surface ions $A_1 = A_2 = 10^{-18}$ m$^2$, $T= 300$ K, other parameters are listed in **Table SI.**

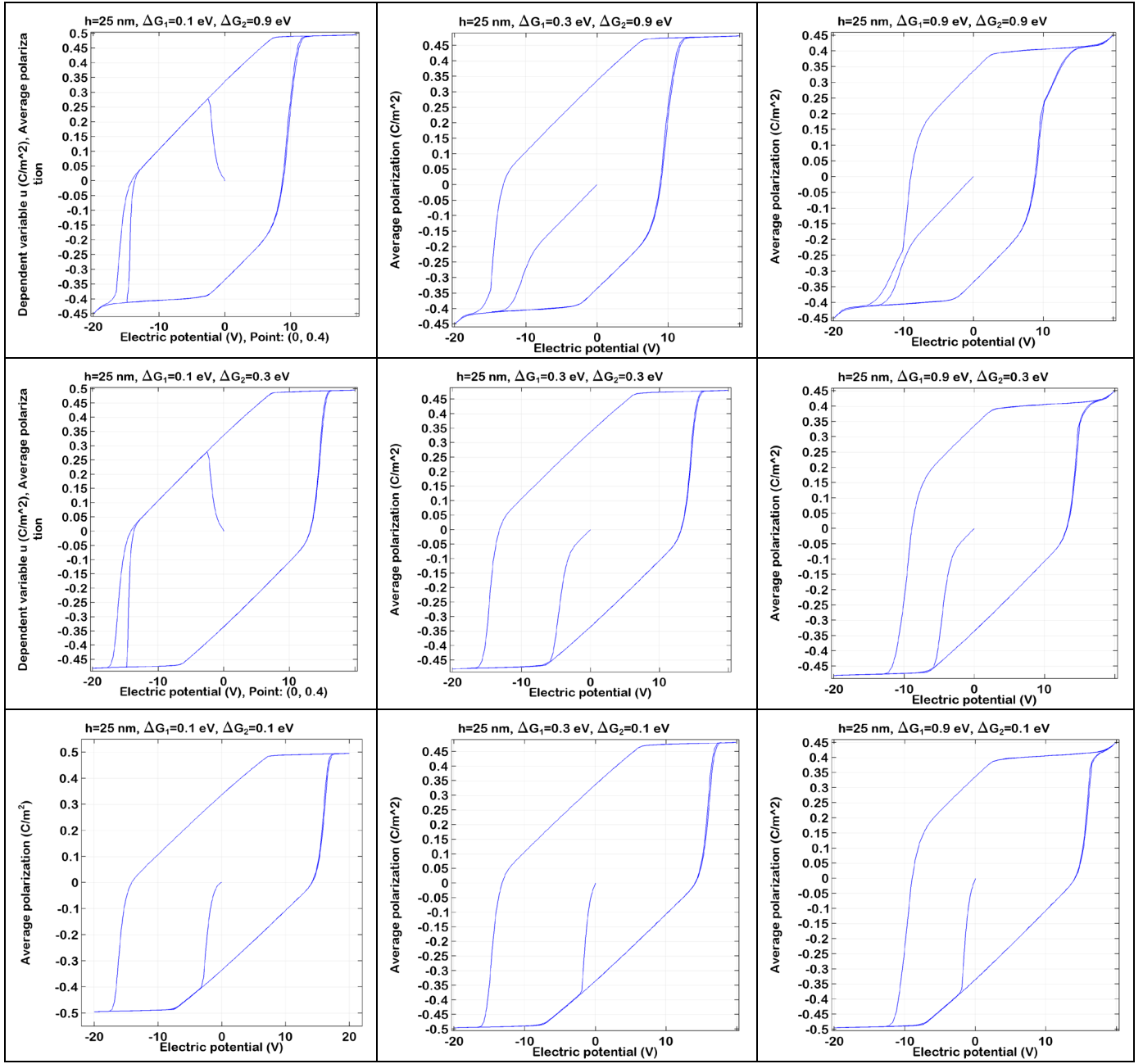

**FIGURE S3.** Dependences $P(U)$ calculated at fixed $\{\Delta G_1^{00}, \Delta G_2^{00}\}$ indicated in the labels for a 25-nm PZT film. Other parameters are the same as in **Fig.S2.**

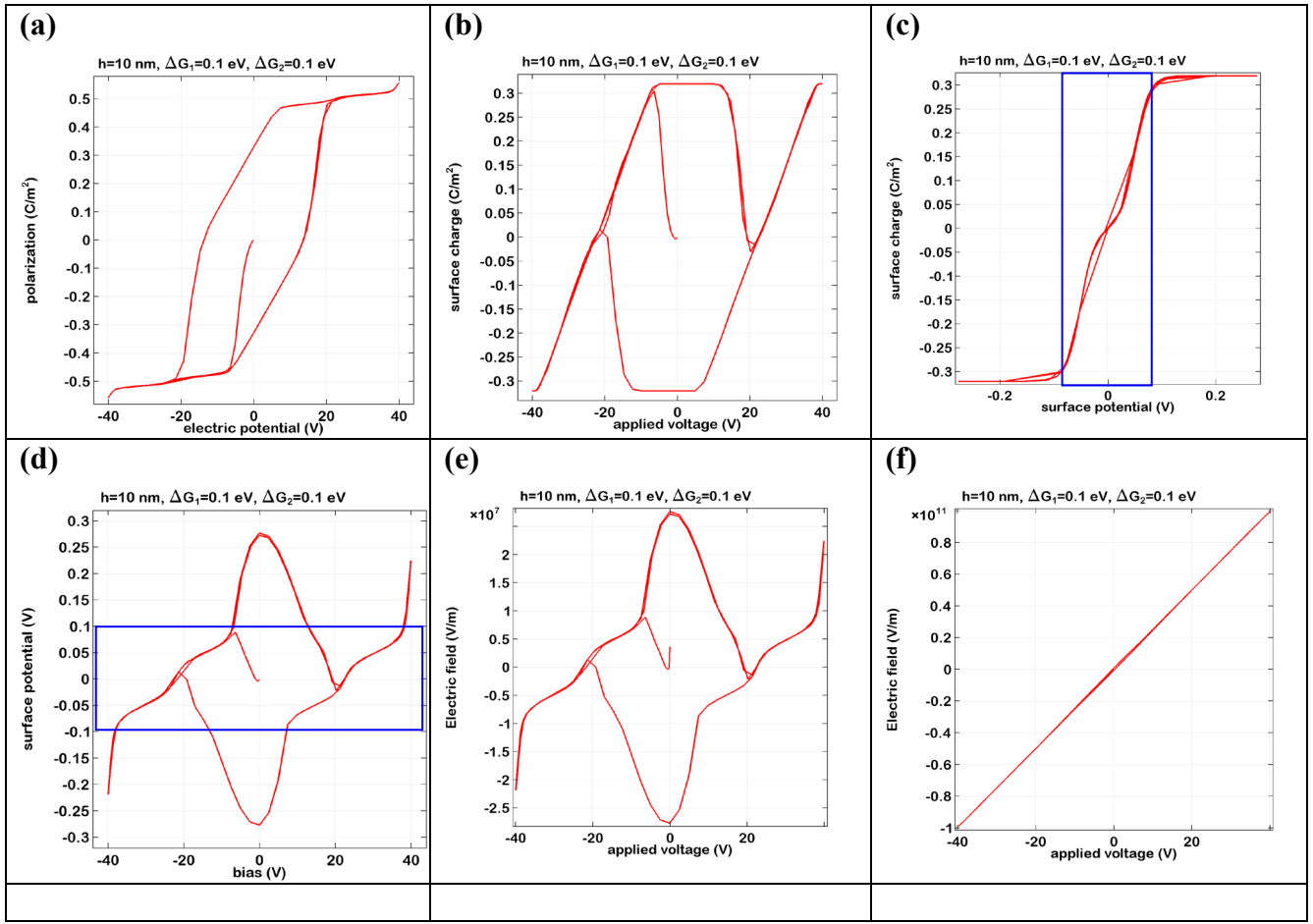

**FIGURE S4.** Dependences of $P(U)$ **(a)**, $\sigma(U)$ **(b)**, $\sigma(\Psi)$ **(c)**, $\varphi(U)$ **(d)** and $E(U)$ in the film **(e)** and in the gap **(f)** vs. the voltage $U$ applied to the top electrode calculated at fixed $\{\Delta G_1^{00}, \Delta G_2^{00}\} = \{0.1, 0.1\} eV$ for a 10-nm PZT film [plots (a)-(d), respectively]. $\varphi(U)$ is calculated at the ferroelectric surface, where the surface ions are located. $E(U)$ is calculated right below (nonlinear, e) and above (linear, f) the ferroelectric surface Other parameters are the same as in **Fig.S2.**

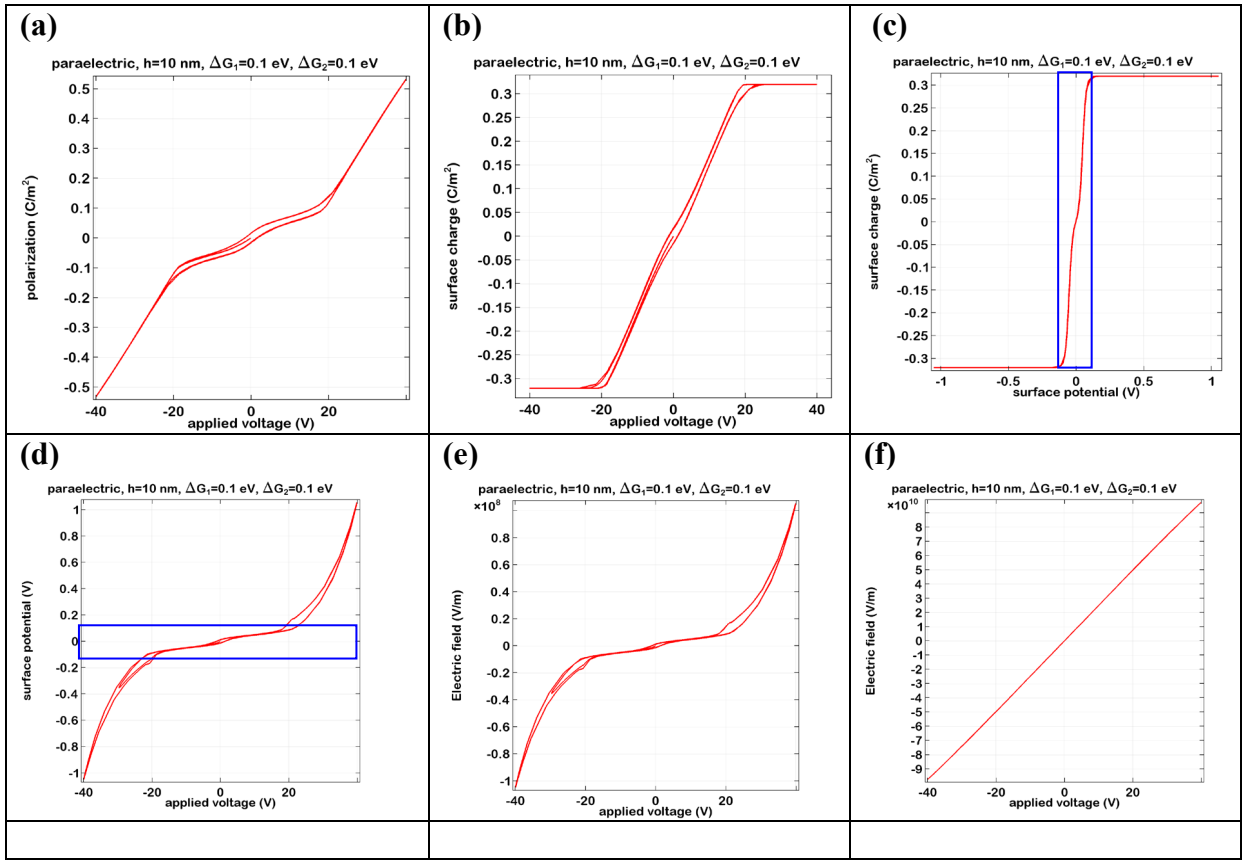

**FIGURE S5.** Dependences of $P(U)$ **(a)**, $\sigma(U)$ **(b)**, $\sigma(\Psi)$ **(c)**, $\varphi(U)$ **(d)** and $E(U)$ in the film **(e)** and in the gap **(f)** vs. the voltage $U$ applied to the top electrode calculated at fixed $\{\Delta G_1^{00}, \Delta G_2^{00}\} = \{0.1, 0.1\} eV$ for a 10-nm paraelectric film [plots (a)-(d), respectively]. $\varphi(U)$ is calculated at the ferroelectric surface, where the surface ions are located. $E(U)$ is calculated right below (nonlinear, e) and above (linear, f) the ferroelectric surface.

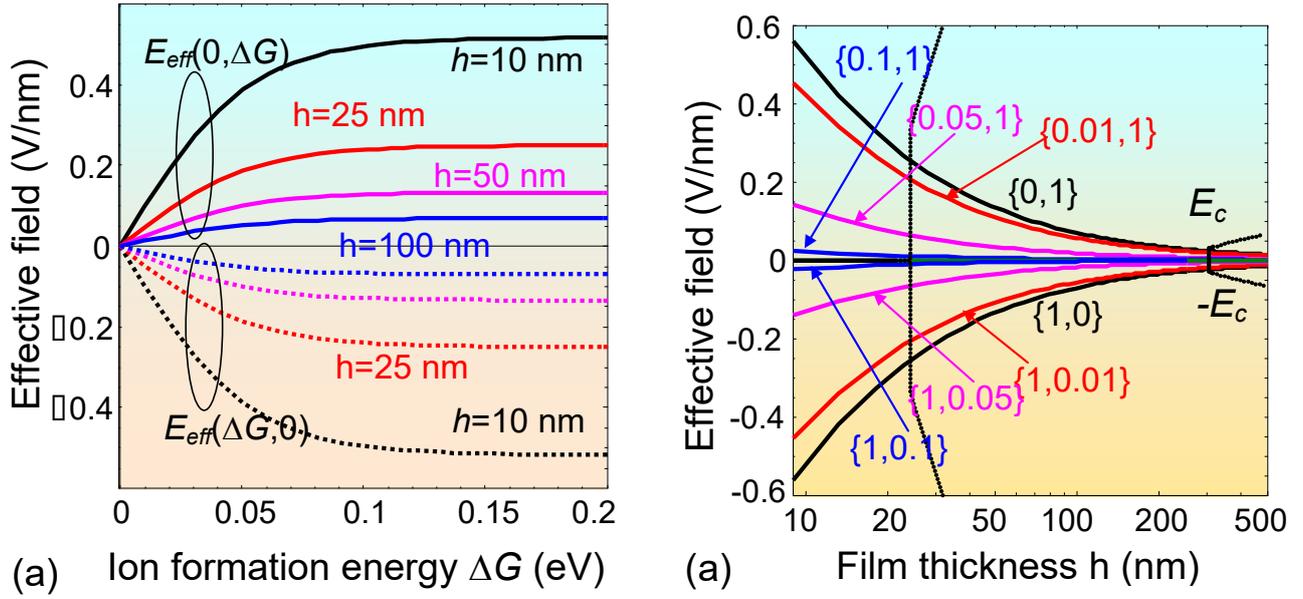

**FIGURE S6.** (a) Effective field $E_0(\Delta G_1^{00}, \Delta G_2^{00})$ induced by surface ions calculated in dependence on the surface ions formation energies $\Delta G_i^{00}$. Different curves correspond to PZT film thickness $h$= 10, 25, 50 and 100 nm (as indicated by legends). Solid curves are $E_0(0, \Delta G_2^{00})$, and dotted curves are $E_0(\Delta G_1^{00}, 0)$. (b) Thickness dependence of effective field $E_0[\Delta G_i^{00}, h]$ calculated for several $\{\Delta G_1^{00}, \Delta G_2^{00}\}$ values (legends near the curves in eV) and PZT parameters. Positive and negative coercive fields $\pm E_C[h]$ are shown by dashed curves. Permittivity of the dielectric gap is $\varepsilon_d$=1, its thickness $\lambda$=0.4 nm and saturation area of the surface ions $A_1 = A_2 = 10^{-18}$ m², $U$=0, $T$= 300 K, other parameters are listed in **Table SI.**

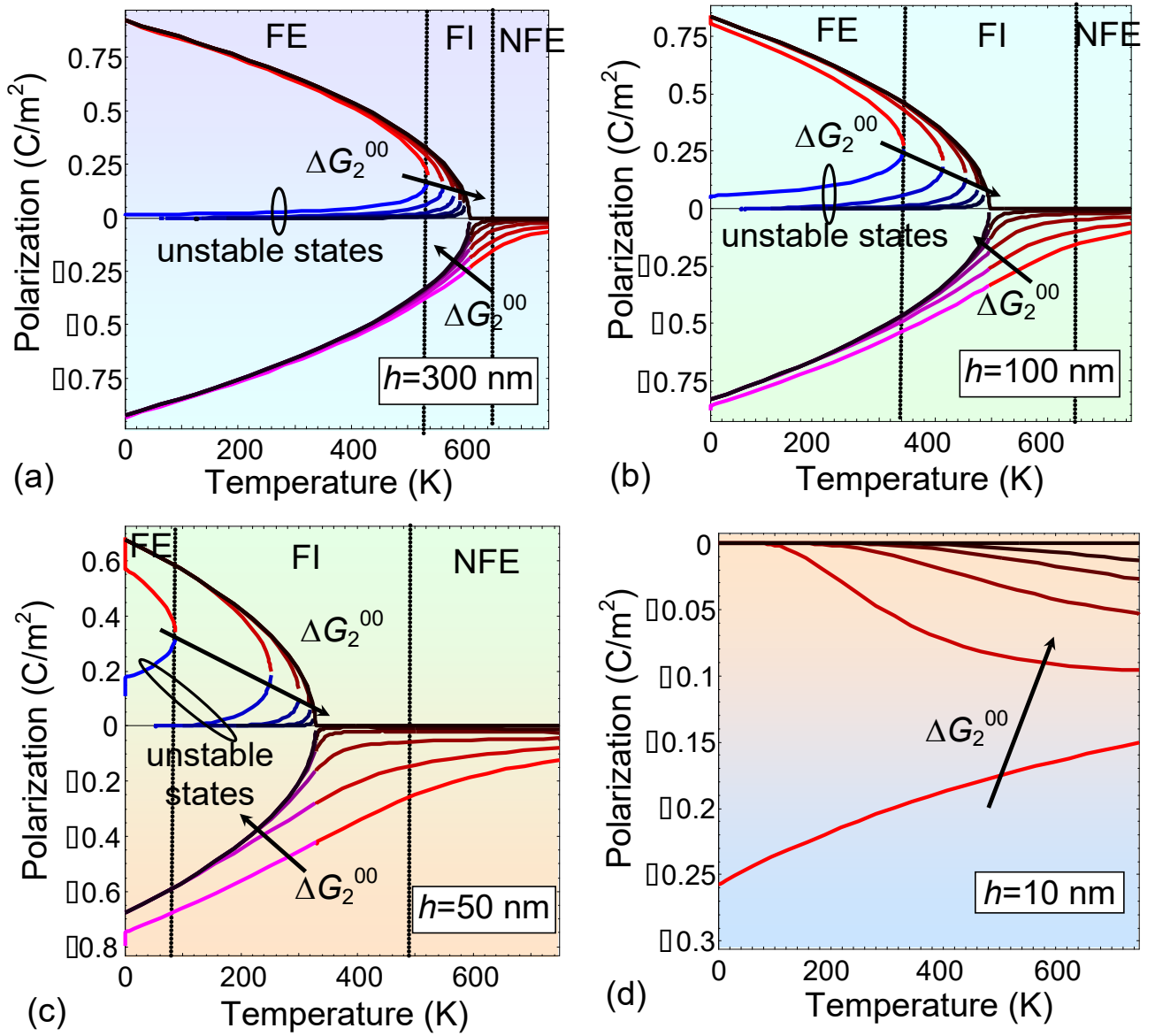

**FIGURE S6.** Temperature dependences of the average polarization calculated in the case of screening by ions, zero applied voltage and different values of film thickness $h$ = 300 nm **(a)**, 100 nm **(b)**, 50 nm **(c)** and 10 nm **(d)**. Each curve in the one-colored group corresponds to the definite value of $\Delta G_2^{00}$ = (0, 0.05, 0.1, 0.15, 0.2, 0.9) eV. Arrow indicates the direction of $\Delta G_2^{00}$ increase. An arrow indicates the direction of $\Delta G_2^{00}$ increase. Unstable branches are shown by blue colors. Permittivity of the dielectric gap is $\varepsilon_d$=10, its thickness $\lambda$=0.4 nm, $\Delta G_1^{00}$ = 1eV, and saturation area of the surface ions $A_1 = A_2 = 10^{-18}$ m$^2$, $U$=0, other parameters are listed in **Table SI.**

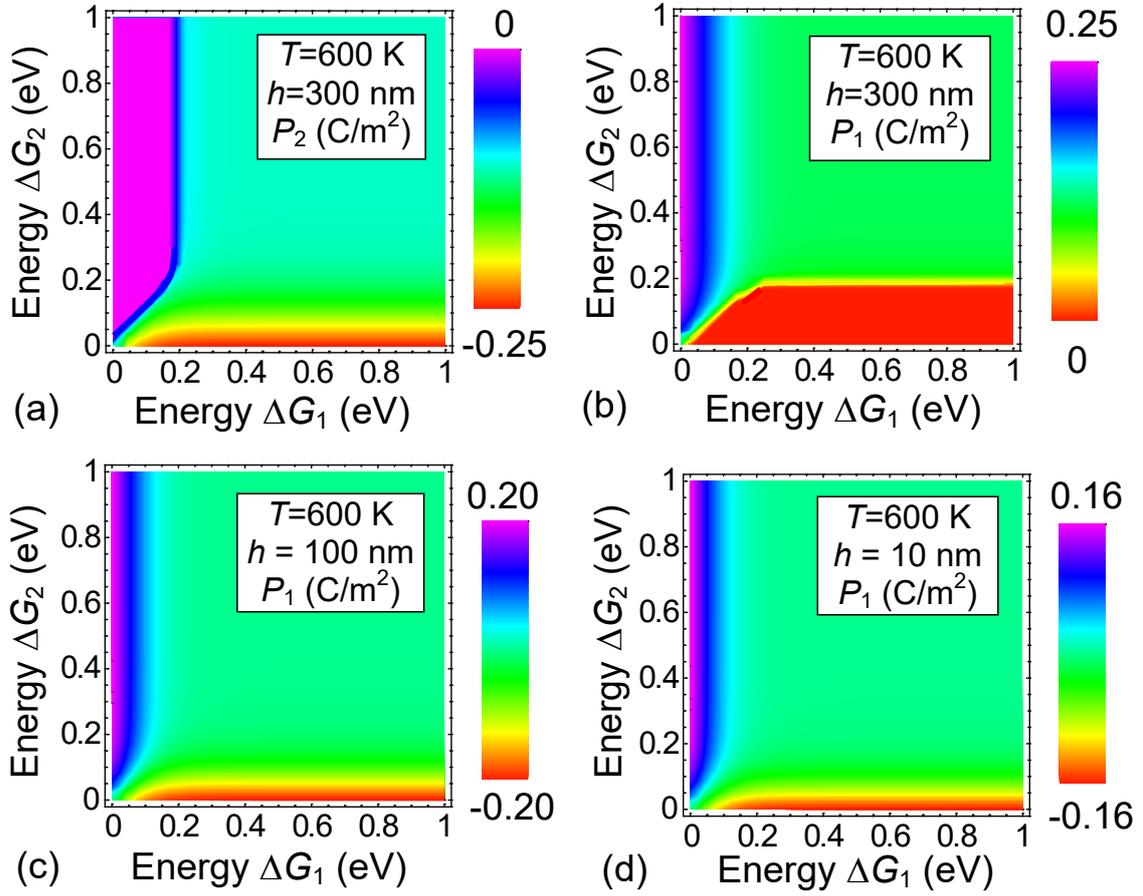

**FIGURE S7.** Dependence of the stable polar states on the surface ions formation energies $\Delta G_i^{00}$ calculated for PZT film thickness $h$= 300, 100 and 10 nm at $T$= 600 K (legends at the curves). Saturation area of the surface ions $A_1 = A_2 = 10^{-18}$ m², dielectric gap thickness $\lambda$=0.4 nm, its permittivity $\varepsilon_d$=10, and $U$=0, and other parameters are listed in **Table SI**. Color scale bar is polarization in C/m². All existing stable and metastable polar states at given thickness and temperature are shown (denoted as the roots P1 and/or P2 of Eq.(14)). Unstable polar sates (denoted as the root P3 of Eq.(14)) are shown in **Fig.S8.**

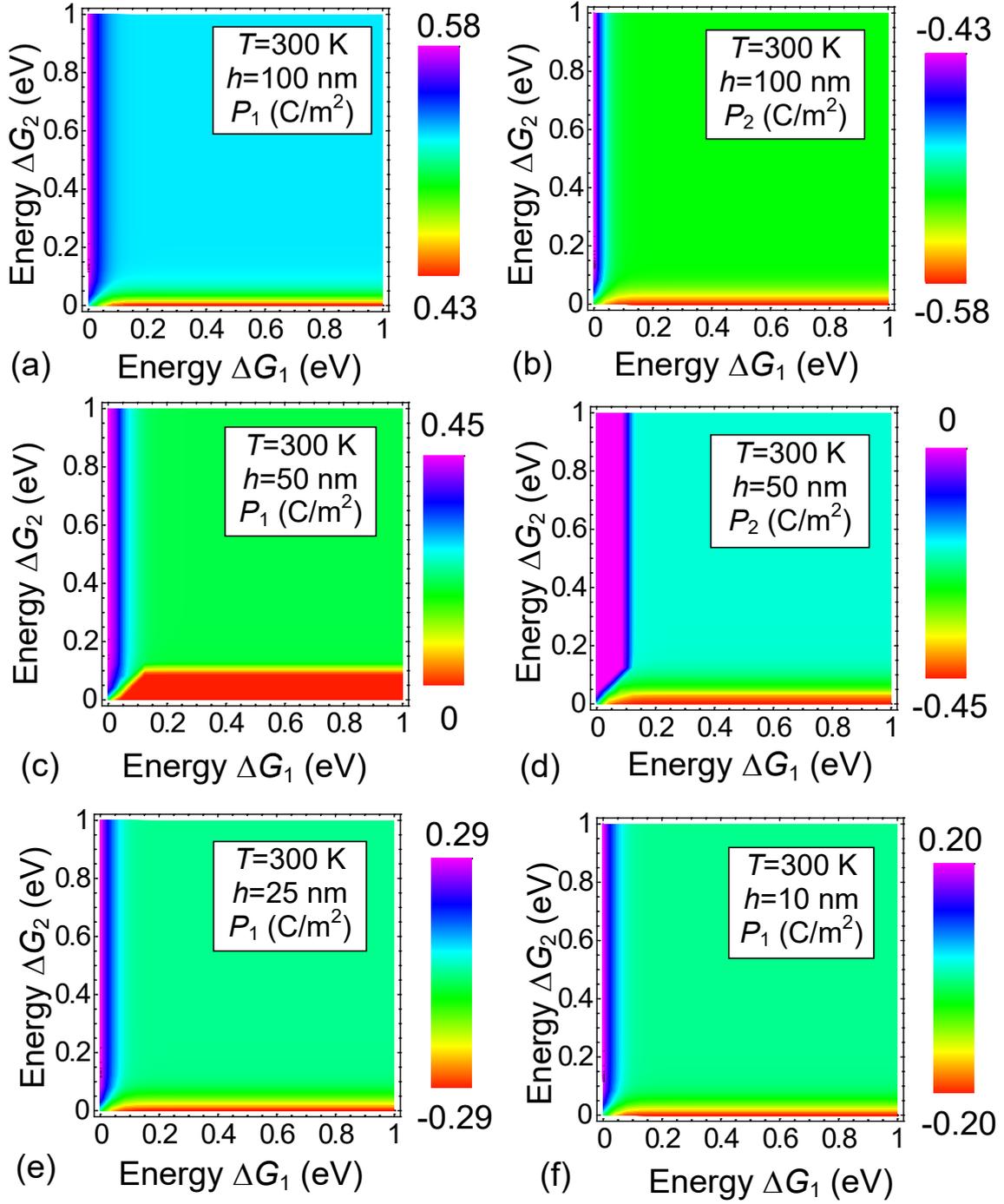

**FIGURE S8.** Dependence of the stable polar states on the surface ions formation energies $\Delta G_i^{00}$ calculated for PZT film thickness $h$= 100, 50, 25 and 10 nm at $T$= 300 K (legends at the curves). Saturation area of the surface ions $A_1 = A_2 = 10^{-18}$ m$^2$, dielectric gap thickness $\lambda$=0.4 nm, its permittivity $\varepsilon_d$=10, and $U$=0, and other parameters are listed in **Table SI**. Color scale bar is polarization in C/m$^2$. All existing stable and metastable polar states at given thickness and temperature are shown.

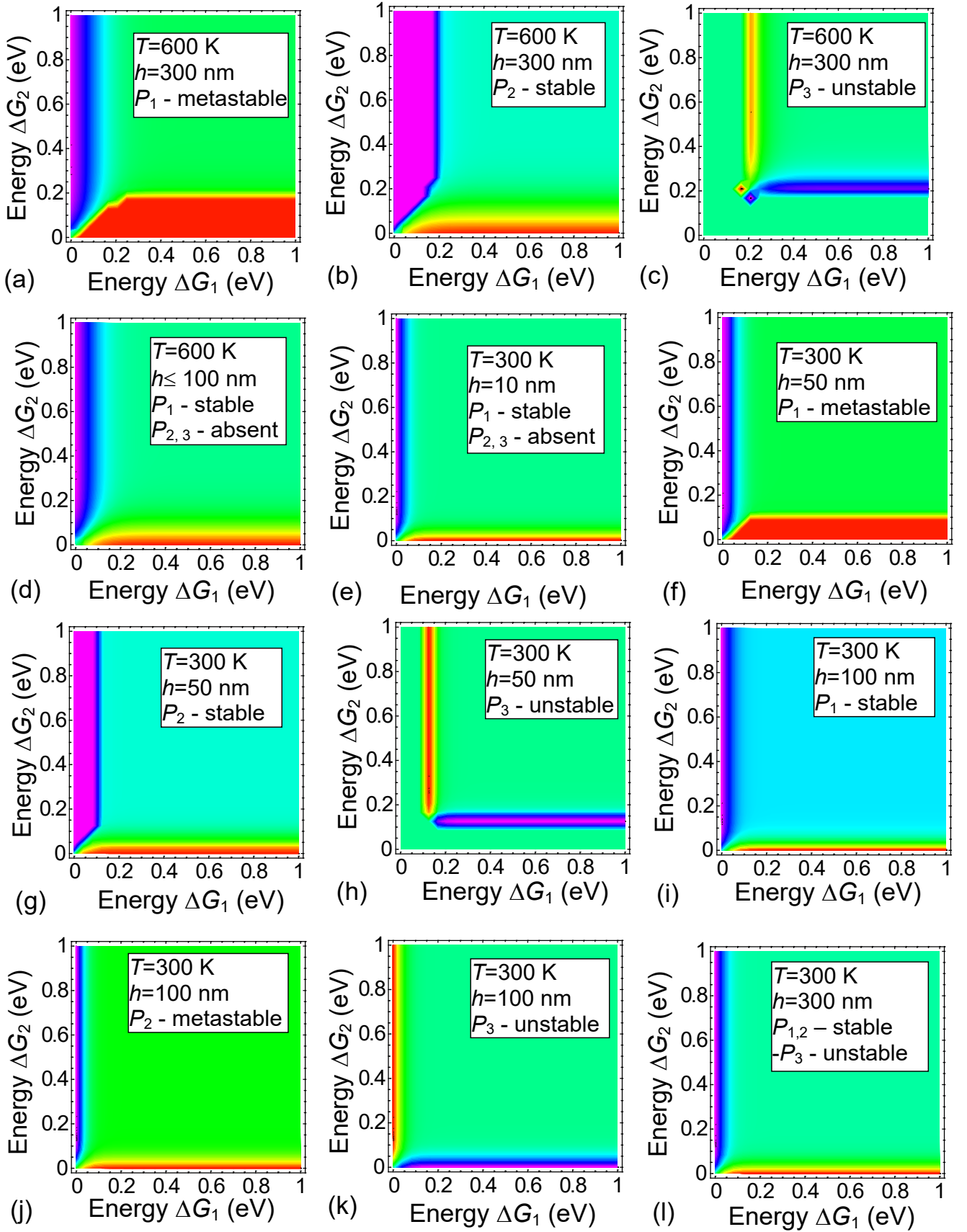

**FIGURE S9.** Stable, metastable and unstable polar states in dependence on the surface ions formation energies $\Delta G_i^{00}$ calculated for PZT film thickness $h$= 10, 50, 100 and 300 nm and $T$= 600, 300 K (legends at the curves). Saturation area of the surface ions $A_1 = A_2 = 10^{-18}$ m², dielectric gap thickness λ=0.4 nm, its permittivity $\varepsilon_d$ =10, and $U$=0, and other parameters are listed in **Table SI**.

12